\documentclass{SciPost}

\hypersetup{
    colorlinks,
    linkcolor={red!50!black},
    citecolor={blue!50!black},
    urlcolor={blue!80!black}
}

\usepackage[bitstream-charter]{mathdesign}
\usepackage{booktabs}
\usepackage[export]{adjustbox}
\usepackage{subcaption}

\usepackage{tikz}
\usetikzlibrary{shapes.geometric, arrows,chains}
\usetikzlibrary{decorations.markings}

\usepackage{tikz-feynman}
\tikzfeynmanset{compat=1.1.0}
\tikzfeynmanset{warn luatex=false}

\usepackage{array}

\DeclareSymbolFont{usualmathcal}{OMS}{cmsy}{m}{n}
\DeclareSymbolFontAlphabet{\mathcal}{usualmathcal}

\fancypagestyle{SPstyle}{
\fancyhf{}
\lhead{\colorbox{scipostblue}{\bf \color{white} ~SciPost Physics Community Reports }}
\rhead{{\bf \color{scipostdeepblue} ~Submission }}

\fancyfoot[C]{\textbf{\thepage}}
}

\newcommand{\POWHEGBOXVV}{{\tt{POWHEG-BOX-V2}}}

\newcommand{\mhh}{m_{HH}}
\newcommand{\pth}{p_{T,H}}

\newcommand{\msbar}{\overline{\rm{MS}}}

\usepackage{comment}

\newcommand{\renscale}{\mu_r}
\newcommand{\facscale}{\mu_f}
 
\begin{document}

\pagestyle{SPstyle}
%
%
%
\begin{center}{\Large \textbf{\color{scipostdeepblue}{
        Investigating bottom quark effects in Higgs boson pair production 
}}}\end{center}

\begin{center}\textbf{
David B\"urg\textsuperscript{1},
Gudrun Heinrich\textsuperscript{1},
Philipp Rendler\textsuperscript{1},
}\end{center}

\begin{center}
{\bf 1} Institute for Theoretical Physics, Karlsruhe Institute
  of Technology (KIT), 76131 Karlsruhe, Germany
%
\\[\baselineskip]
\href{mailto:david.buerg@student.kit.edu}{\small david.buerg@student.kit.edu},
\href{mailto:gudrun.heinrich@kit.edu}{\small gudrun.heinrich@kit.edu}, 
\href{mailto:philipp.rendler}{\small philipp.rendler@kit.edu}
\end{center}

\section*{\color{scipostdeepblue}{Abstract}}
\textbf{\boldmath{%
We investigate different effects of bottom quark masses in Higgs boson pair production, considering both bottom quark loops and bottom quarks in the initial state. While the effects are negligible at total cross section level, the situation is different for the Higgs boson pair invariant mass distribution close to the production threshold of the two Higgs bosons.
%
}}

\vspace{\baselineskip}

\noindent\textcolor{white!90!black}{%
\fbox{\parbox{0.975\linewidth}{%
\textcolor{white!40!black}{\begin{tabular}{lr}%
  \begin{minipage}{0.6\textwidth}%
    {\small Copyright attribution to authors. \newline
    This work is a submission to SciPost Phys. Comm. Rep. \newline
    License information to appear upon publication. \newline
    Publication information to appear upon publication.}
  \end{minipage} & \begin{minipage}{0.4\textwidth}
    {\small Received Date \newline Accepted Date \newline Published Date}%
  \end{minipage}
\end{tabular}}
}}
}\newpage


\vspace{10pt}
\noindent\rule{\textwidth}{1pt}
\tableofcontents
\noindent\rule{\textwidth}{1pt}
\vspace{10pt}


\section{Introduction}
Higgs boson pair production is the prime process to constrain the trilinear Higgs-boson self-coupling, and the observation of di-Higgs production is an exciting prospect for the High-Luminosity LHC~\cite{ATLAS:2025eii}.

Higgs boson pairs at the LHC are dominantly produced through gluon fusion via top quark loops, due to the large gluon luminosity at the LHC and due to the large Yukawa coupling of the Higgs boson to top quarks.
The effects of bottom quark loops in the gluon fusion channel  are usually neglected, as they contribute at permille level to the total cross section for $gg\to HH$ if Standard Model (SM) couplings are assumed. Similar considerations hold for bottom quarks in the initial state.
However, these considerations are based on studies at the level of total cross sections, while at differential level, for example considering the Higgs boson pair invariant mass distribution, bottom quark effects can play a non-negligible role, in particular in view of the prospects for high-statistics data at the High-Luminosity LHC.
Furthermore, bottom quarks are usually treated in the $\msbar$ scheme, which introduces a dependence on an additional unphysical scale, the mass renormalisation scale $\mu_m$. The dependence on $\mu_m$ so far only has been investigated for top quark loops~\cite{Baglio:2018lrj,Bagnaschi:2023rbx,Jaskiewicz:2024xkd,Davies:2025ghl,Davies:2025qjr,Chen:2026zmi}, but not for bottom quark loops. 
Therefore, we perform a detailed analysis of the effects of bottom quark loops in the on-shell and $\msbar$ schemes, comparing also the four- and five-flavour schemes, and investigating the influence of the order of the $\alpha_s$-running when performing the mass conversion. 
For single Higgs production, a detailed comparison of renormalisation- and flavour-schemes has been performed in Ref.~\cite{Czakon:2024ywb} up to NNLO, showing that the difference between the four- and five flavour schemes are small, even at lower orders.
For di-Higgs production, 
 we limit ourselves to $gg\to HH$ at one loop to obtain a first estimate of these effects. The largest effects of the bottom quarks will of course come from the top-bottom interference contribution. 
 The full two-loop corrections involving bottom quark loops are not yet known; however, the planar master integrals in a small-mass expansion have been presented in Ref.~\cite{Hu:2025aeo}. 

Secondly, we consider Higgs boson pair production in bottom quark annihilation at next-to-leading order (NLO). Assuming a massless bottom-antibottom quark pair in the initial state while retaining a non-zero Yukawa coupling allows for the production of two Higgs bosons already at tree level. The contribution of this channel to the total cross section has been studied up to NNLO in Refs.~\cite{Dawson:2006dm, AH:2018tzg, Ahmed:2021hrf}. For a proton-proton centre-of-mass energy of $14\,\mathrm{TeV}$, the resulting total cross section was found to be of the order of $\mathcal{O}(0.03\,\mathrm{fb})$, depending on the exact choice of renormalisation and factorisation scales. Consequently, its impact on the total cross section is negligible compared to the gluon fusion channel. 

However, the small contribution at the level of the total cross section does not preclude sizeable effects in differential observables. In this paper, we investigate the impact of bottom quark annihilation on differential distributions. We observe that including this channel at NLO QCD results in relative corrections of up to $+18\%$ in the invariant mass distribution compared to LO gluon fusion. 

In Section~\ref{sec:gluon_fusion}, we investigate the effect of different scheme choices on the inclusion of bottom quarks at LO in gluon fusion. Section~\ref{sec:initial_bottom} is dedicated to Higgs boson pair production in bottom quark annihilation, before we conclude in Section~\ref{sec:conclusion}.

\section{Effects of bottom quark loops in the on-shell and $\msbar$ schemes}
\label{sec:gluon_fusion}
In this section, we investigate the effects of different mass renormalisation schemes for both the top- and the bottom quark, on the leading order cross section of Higgs boson pair production via gluon fusion in proton-proton scattering. The renormalisation schemes considered are the on-shell and the $\overline{\mathrm{MS}}$ mass renormalisation schemes.

\subsection{Framework of the calculation}

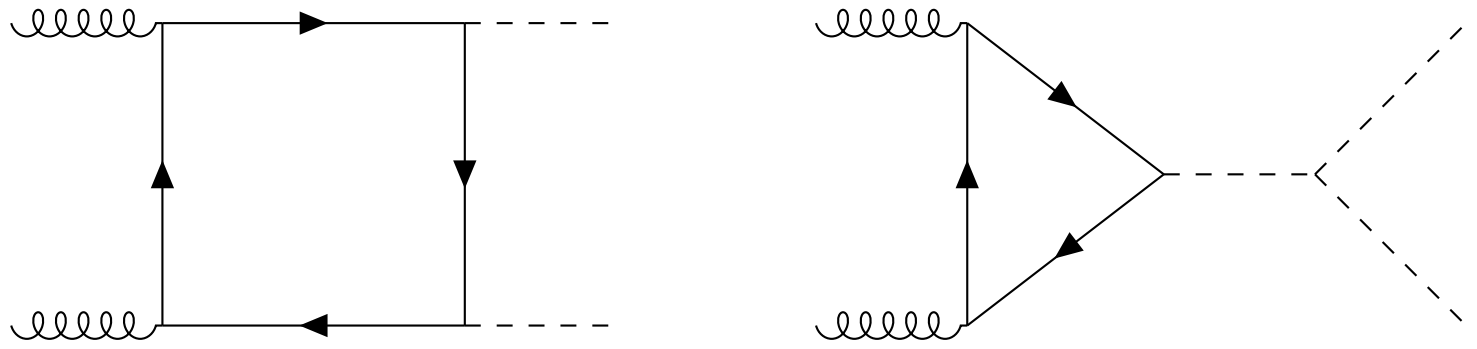
\begin{figure}[tb]
\centering 
\begin{tikzpicture}
    \begin{feynman}[inline = (hi)]
        \vertex (hi);
        \vertex [right = 1.0cm of hi] (ha);
        \vertex [right = 2.0cm of ha] (hb);
        \vertex [right = 1.0cm of hb] (hf);
        \vertex [above = 1.0cm of hi] (i1);
        \vertex [below = 1.0cm of hi] (i2);
        \vertex [above = 1.0cm of ha] (a1);
        \vertex [below = 1.0cm of ha] (a2);
        \vertex [above = 1.0cm of hb] (b1);
        \vertex [below = 1.0cm of hb] (b2);
        \vertex [above = 1.0cm of hf] (f1);
        \vertex [below = 1.0cm of hf] (f2);

        \diagram*{
            (i1) -- [gluon] (a1), 
            (i2) -- [gluon] (a2),  
            (a1) -- [fermion] (b1) -- [fermion] (b2) -- [fermion] (a2) -- [fermion] (a1),
            (b1) -- [scalar] (f1), 
            (b2) -- [scalar] (f2)
        };
    \end{feynman}
\end{tikzpicture} 
\hspace{1cm}
\begin{tikzpicture}
    \begin{feynman}[inline = (hi)]
        \vertex (hi);
        \vertex [right = 1.0cm of hi] (ha);
        \vertex [right = 1.3cm of ha] (hb);
        \vertex [right = 1.0cm of hb] (hc);
        \vertex [right = 1.0cm of hc] (hf);
        \vertex [above = 1.0cm of hi] (i1);
        \vertex [below = 1.0cm of hi] (i2);
        \vertex [above = 1.0cm of ha] (a1);
        \vertex [below = 1.0cm of ha] (a2);
        \vertex [above = 1.0cm of hf] (f1);
        \vertex [below = 1.0cm of hf] (f2);

        \diagram*{
            (i1) -- [gluon] (a1), 
            (i2) -- [gluon] (a2),  
            (a1) -- [fermion] (hb) -- [fermion] (a2) -- [fermion] (a1),
            (hb) -- [scalar] (hc) -- [scalar] (f1), 
            (hc) -- [scalar] (f2)
        };
    \end{feynman}
\end{tikzpicture} 
\caption{Example diagrams for Higgs boson pair production in gluon fusion, where we consider both top- and bottom quark loops.}
\label{fig:diags}
\end{figure}

Higgs boson pair production via gluon fusion, $g(p_1)g(p_2) \rightarrow H(p_3)H(p_4)$, does not occur at the tree-level in the SM and therefore is mediated at leading order via quark loops. The contributing diagrams are shown in Figure~\ref{fig:diags}. The matrix element for the partonic scattering process has been analytically computed in Ref.~\cite{Glover:1987nx}. Denoting the polarization vectors of the incoming gluons by $\epsilon^{(\lambda_1)}_\mu(p_1)$ and $\epsilon^{(\lambda_2)}_\nu(p_2)$, the matrix element can be written as
\begin{equation}
    \mathcal{M} = \epsilon^{(\lambda_1)}_\mu(p_1) \epsilon^{(\lambda_2)}_\nu(p_2) \mathcal{M}^{\mu\nu} ~,
\end{equation}
where $\mathcal{M}^{\mu\nu}$ can be expressed as
\begin{equation}
    \mathcal{M}^{\mu\nu} = \frac{G_F \alpha_s \delta^{ab}}{4 \sqrt{2} \pi} \left(F_1(\hat{s}, \hat{t}, m_H^2, m_f^2) T_1^{\mu\nu} + F_2(\hat{s}, \hat{t}, m_H^2, m_f^2) T_2^{\mu\nu}\right)
\end{equation}
by introducing the Lorentz tensors
\begin{align}
    T_1^{\mu\nu} & = g^{\mu\nu} - \frac{p_1^\nu p_2^\mu}{p_1 \cdot p_2}~,\\
    T_2^{\mu\nu} & = g^{\mu\nu} + \frac{1}{p_T^2 (p_1 \cdot p_2)} \left(m_H^2 p_1^\nu p_2^\mu - 2 (p_1 \cdot p_3) p_2^\mu p_3^\nu -2 (p_2 \cdot p_3)  p_1^\nu p_3^\mu + 2 (p_1 \cdot p_2) p_3^\mu p_3^\nu \right)
\end{align}
and the form factors $F_1$ and $F_2$, which can be split into triangle and box contributions
\begin{align}
    F_1(\hat{s}, \hat{t}, m_H^2, m_f^2) & = F_{1\triangle}(\hat{s}, \hat{t}, m_H^2, m_f^2) + F_{1\Box}(\hat{s}, \hat{t}, m_H^2, m_f^2)~, \\
    F_2(\hat{s}, \hat{t}, m_H^2, m_f^2) & = F_{2\Box}(\hat{s}, \hat{t}, m_H^2, m_f^2) ~.
\end{align}
The partonic cross section is then obtained by integrating over the phase space and taking the average or sum over the quantum numbers of the incoming and outgoing particles, respectively, leading to
\begin{equation}
    \hat{\sigma}(gg \rightarrow HH) = \frac{G_F^2 \alpha_s^2}{2^{14} \pi^3 \hat{s}} \int_{\hat{t}_-}^{\hat{t}_+} \mathrm{d}\hat{t} \left( |F_1(\hat{s}, \hat{t}, m_H^2, m_f^2)|^2 + |F_2(\hat{s}, \hat{t}, m_H^2, m_f^2)|^2\right) 
\end{equation}
with
\begin{equation}
    \hat{t}_\pm = m_H^2 -\frac{\hat{s}}{2} \left(1 \mp \sqrt{1 - \frac{4 m_H^2}{\hat{s}}}\right) ~.
\end{equation}
To obtain the leading order cross section of Higgs boson pair production via gluon fusion in proton-proton scattering, $pp \rightarrow gg \rightarrow HH$, the partonic cross section must be integrated over the parton distribution function for the gluons $f_g(x_i,\facscale)$ by
\begin{equation}
    \sigma(pp \rightarrow gg \rightarrow HH) = \int_0^1 \mathrm{d}x_1 \int_0^1 \mathrm{d}x_2 \; f_g(x_1, \facscale) f_g(x_2, \facscale) \hat{\sigma}(gg \rightarrow HH) \;.
\end{equation}
This integration is performed using the \texttt{POWHEG-BOX-V2} framework \cite{Nason:2004rx, Frixione:2007vw, Alioli:2010xd} with \texttt{LHAPDF6} \cite{Buckley:2014ana} to include either the parton distribution function \texttt{PDF4LHC21\_40} or \texttt{PDF4LHC21\_40\_nf4} \cite{PDF4LHCWorkingGroup:2022cjn}, depending on the used fixed flavour number scheme. The calculation is based on an extension of the NLO QCD program \texttt{ggHH}~\cite{Heinrich:2017kxx,Heinrich:2020ckp}, in which the leading order partonic cross section is implemented analytically.

\subsubsection{$\overline{\mathrm{MS}}$ scheme}
To relate the on-shell and the $\overline{\mathrm{MS}}$ mass renormalisation schemes to each other, we write the bare quark $m_0$ as a product of the mass renormalisation factor and the renormalised mass for both schemes
\begin{equation}
    m_0 = Z^\mathrm{OS} m^\mathrm{OS} = Z^{\overline{\mathrm{MS}}}(\mu) m^{\overline{\mathrm{MS}}}(\mu) ~.
\end{equation}
The ratio between the on-shell mass $m^\mathrm{OS}$ and the $\overline{\mathrm{MS}}$ mass $m^{\overline{\mathrm{MS}}}(\mu)$ is then given by the ratio of the mass renormalisation factors
\begin{equation}
    \frac{ m^{\overline{\mathrm{MS}}} }{ m^\mathrm{OS} } = \frac{ Z^\mathrm{OS} }{ Z^{\overline{\mathrm{MS}}} } ~,
\end{equation}
which has been computed in Ref.~\cite{Melnikov:2000qh} up to $\mathcal{O}(\alpha_s^3(\mu))$ assuming that the masses of the light quarks are negligible compared to those of the heavy quarks.\newline

For a theory with gauge group $\mathrm{SU}(3)$ such as QCD and $N_F = N_L + N_H$ active quark flavours, where $N_L$ denotes the number of light quark flavours and $N_H$ the number of heavy quark flavours with on-shell mass $m^\mathrm{OS}$, this relation between the masses of the heavy quarks is given by
\begin{equation}
\label{eq:conversion_formula_OS_MSbar_mass}
    \frac{m^{\overline{\mathrm{MS}}}(m^\mathrm{OS})}{m^\mathrm{OS}} = 1 + b^{(1)} \frac{\alpha_s(m^\mathrm{OS})}{\pi} + b^{(2)} \left(\frac{\alpha_s(m^\mathrm{OS})}{\pi}\right)^2 + b^{(3)} \left(\frac{\alpha_s(m^\mathrm{OS})}{\pi}\right)^3 + \mathcal{O}\left(\alpha_s^4(m^\mathrm{OS})\right)
\end{equation}
with the strong coupling constant $\alpha_s(m^\mathrm{OS}) = \alpha_s^{(N_L+N_H)}(m^\mathrm{OS})$ and the coefficients:
\begin{align}
    b^{(1)} = & -\frac{4}{3}~, \nonumber \\
    b^{(2)} = & -\frac{3305}{288} - \frac{4}{3}\zeta_2 + \frac{1}{6}\zeta_3 - \frac{2}{3}\zeta_2\ln(2) + \left(\frac{71}{144} + \frac{1}{3}\zeta_2\right)N_L + \left(\frac{143}{144} - \frac{2}{3}\zeta_2\right)N_H ~, \nonumber \\
    b^{(3)} = &  -\frac{1259285}{10368} + \frac{76}{9}a_4 - \frac{24995}{432}\zeta_2 + \frac{73}{72}\zeta_3 + \frac{1331}{72}\zeta_2\zeta_3 + \frac{1705}{144}\zeta_4 - \frac{1705}{216}\zeta_5 \nonumber \\
    & - \frac{77}{3}\zeta_2\ln(2) + \frac{14}{9}\zeta_2\ln^2(2) + \frac{19}{54}\ln^4(2) + \left(\frac{86159}{7776} - \frac{8}{27}a_4 + \frac{941}{108}\zeta_2 + \frac{707}{216}\zeta_3 \right. \nonumber \\
    & \left. -\frac{305}{108}\zeta_4 + \frac{22}{27}\zeta_2\ln(2) - \frac{4}{27}\zeta_2\ln^2(2) - \frac{1}{81}\ln^4(2)\right)N_L + \left(\frac{157763}{7776} - \frac{8}{27}a_4 \right. \nonumber \\
    & \left. - \frac{13483}{324}\zeta_2 - \frac{751}{216}\zeta_3 + \frac{3}{2}\zeta_2\zeta_3 - \frac{205}{54}\zeta_4 - \frac{5}{4}\zeta_5 + \frac{1280}{27}\zeta_2\ln(2) + \frac{2}{27}\zeta_2\ln^2(2) \right. \nonumber \\
    & \left. - \frac{1}{81}\ln^4(2) \right)N_H + \left(-\frac{2353}{23328} - \frac{13}{54}\zeta_2 - \frac{7}{54}\zeta_3 \right)N_L^2 + \left(-\frac{5917}{11664} + \frac{13}{54}\zeta_2 \right.\nonumber \\
    & \left. + \frac{2}{27}\zeta_3 \right)N_L N_H + \left(-\frac{9481}{23328} + \frac{8}{135}\zeta_2 + \frac{11}{54}\zeta_3\right)N_H^2 ~,
\end{align}
where $a_4$ denotes the polylogarithm $\mathrm{Li}_4(1/2) = \sum_{n=1}^\infty \frac{1}{4^n} \left(\frac{1}{2}\right)^n \simeq 0.517479$ and $\zeta_z$ the Riemann zeta function $\zeta(z)$.\newline

We introduce the scale-invariant mass $m^\mathrm{inv}$ through the defining equation
\begin{equation}
    \label{eq:scale_invariant_mass_defining_eq}
    m^\mathrm{inv} = m^{\overline{\mathrm{MS}}}(m^\mathrm{inv})
\end{equation}
and relate this scale-invariant mass $m^\mathrm{inv}$ to the on-shell mass $m^\mathrm{OS}$ for the heavy quark flavour. To do this, we use the $\beta$-function, which satisfies the renormalisation group equation
\begin{equation}
    \label{eq:beta_function_RGEq}
    \mu^2 \frac{\mathrm{d a}}{\mathrm{d}\mu^2} = \beta(a)~.
\end{equation}
Together with the series expansion in the coupling $a(\mu) = \frac{\alpha_s(\mu)}{\pi}$,
\begin{equation}
    \label{eq:beta_function_series_expansion}
    \beta(a) = - \sum_{n=0}^\infty \beta_n a^{n+2} ~,
\end{equation}
the renormalisation group equation of the $\beta$-function yields a series expansion for the coupling 
\begin{align}
\label{eq:couplant_series_expansion}
    \frac{a(\mu_2)}{a(\mu_1)} = & 1 - \beta_0\ln\left(\frac{\mu_2^2}{\mu_1^2}\right) a(\mu_1) + \left(-\beta_1\ln\left(\frac{\mu_2^2}{\mu_1^2}\right) + \beta_0^2\ln^2\left(\frac{\mu_2^2}{\mu_1^2}\right)\right) a^2(\mu_1) \nonumber \\
    & + \left(- \beta_2\ln\left(\frac{\mu_2^2}{\mu_1^2}\right) + \frac{5}{2}\beta_1\beta_0\ln^2\left(\frac{\mu_2^2}{\mu_1^2}\right) - \beta_0^3\ln^3\left(\frac{\mu_2^2}{\mu_1^2}\right) \right) a^3(\mu_1) + \mathcal{O}\left(a^4(\mu_1)\right) ~.
\end{align}
Similarly, we make use of the mass anomalous dimension $\gamma_m$ satisfying the renormalisation group equation
\begin{equation}
    \label{eq:mass_anomalous_dimension_RGEq}
    \gamma_m(a) = \frac{\mu^2}{m^{\overline{\mathrm{MS}}}} \frac{\mathrm{d} m^{\overline{\mathrm{MS}}}}{\mathrm{d} \mu^2} ~.
\end{equation}
In combination with the renormalisation group equation of the $\beta$-function in Eq.~\eqref{eq:beta_function_RGEq}, the series expansion of the $\beta$-function in Eq.~\eqref{eq:beta_function_series_expansion} and the series expansion of the mass anomalous dimension
\begin{equation}
    \gamma_m(a) = - \sum_{n=1}^\infty \gamma_n a^{n+1} ~,
\end{equation}
the renormalisation group equation of the mass anomalous dimension yields the series expansion
\begin{align}
\label{eq:msbar_mass_series_expansion}
    \frac{m^{\overline{\mathrm{MS}}}(\mu_2)}{m^{\overline{\mathrm{MS}}}(\mu_1)} = & 1 - \gamma_0\ln\left(\frac{\mu_2^2}{\mu_1^2}\right) a(\mu_2) + \left(-\gamma_1\ln\left(\frac{\mu_2^2}{\mu_1^2}\right) + \frac{\gamma_0(\gamma_0 - \beta_0)}{2}\ln^2\left(\frac{\mu_2^2}{\mu_1^2}\right)\right) a^2(\mu_2) \nonumber \\
    & + \left(-\gamma_2\ln\left(\frac{\mu_2^2}{\mu_1^2}\right) + \left( \gamma_1\gamma_0 - \gamma_1\beta_0 - \frac{\gamma_0\beta_1}{2}\right)\ln^2\left(\frac{\mu_2^2}{\mu_1^2}\right) + \right. \nonumber \\
    & + \left. \frac{-\gamma_0^3 + 3\gamma_0^2\beta_0 - 2\gamma_0\beta_0^2}{6}\ln^3\left(\frac{\mu_2^2}{\mu_1^2}\right)\right) a^3(\mu_2) + \mathcal{O}\left(a^4(\mu_2)\right) ~,
\end{align}
which describes the running of the $\overline{\mathrm{MS}}$ mass as a function of the renormalisation scale.\newline

Finally, we put together Eqs.~\eqref{eq:couplant_series_expansion} and \eqref{eq:msbar_mass_series_expansion}, the coefficients up to three-loop order of both the QCD $\beta$-function~\cite{Gross:1973id, Politzer:1973fx, Jones:1974mm, Caswell:1974gg, Tarasov:1980au}
\begin{align}
    \beta_0 = & \frac{1}{4} \left(11 - \frac{2}{3} N_F\right) ~, \nonumber \\
    \beta_1 = & \frac{1}{4^2} \left(102 - \frac{38}{3}N_F\right) ~, \nonumber \\
    \beta_2 = & \frac{1}{4^3} \left(\frac{2857}{2} - \frac{5033}{18}N_F + \frac{325}{54}N_F^2\right)
\end{align}
and the QCD mass anomalous dimension~\cite{Tarrach:1980up, Nachtmann:1981zg, Tarasov:1980au, Larin:1993tq}
\begin{align}
\gamma_0 = & 1~, \nonumber \\
    \gamma_1 = & \frac{1}{4^2} \left(\frac{202}{3} - \frac{20}{9}N_F\right)~, \nonumber \\
    \gamma_2 = & \frac{1}{4^3} \left(1249 + \left(-\frac{2216}{27} - \frac{160}{3}\zeta_3\right)N_F - \frac{140}{81}N_F^2\right)
\end{align}
for a theory with $N_F = N_L + N_H$ active quark flavours, as well as Eq.~\eqref{eq:conversion_formula_OS_MSbar_mass} relating the on-shell to the $\overline{\mathrm{MS}}$ mass, to obtain the following relations between the on-shell and the scale-invariant mass of the heavy quark:
\begin{align}
\label{eq:scale_invariant_mass_series_expansion}
    \frac{m^\mathrm{inv}}{m^\mathrm{OS}} = 1 + c^{(1)} a(m^\mathrm{OS}) + c^{(2)} a^2(m^\mathrm{OS}) + c^{(3)} a^3(m^\mathrm{OS}) + \mathcal{O}\left(a^4(m^\mathrm{OS})\right)
\end{align}
with the coefficients
\begin{align}
    c^{(1)} = & b^{(1)} ~, \nonumber \\
    c^{(2)} = & b^{(2)} + \frac{8}{3} ~, \nonumber \\
    c^{(3)} = & b^{(3)} + \frac{433}{16} + \frac{8}{3}\zeta_2 - \frac{1}{3}\zeta_3 + \frac{4}{3}\zeta_2\ln(2) + \left(-\frac{293}{216} - \frac{2}{3}\zeta_2\right)N_L \nonumber \\
    & + \left(-\frac{509}{216} + \frac{4}{3}\zeta_2\right)N_H~,
\end{align}
as well as the inverse equation
\begin{align}
\label{eq:on_shell_mass_series_expansion}
    \frac{m^\mathrm{OS}}{m^\mathrm{inv}} = 1 + d^{(1)} a(m^\mathrm{inv})+ d^{(2)} a^2(m^\mathrm{inv}) + d^{(3)} a^3(m^\mathrm{inv}) + \mathcal{O}\left(a^4(m^\mathrm{inv})\right)
\end{align}
with the coefficients
\begin{align}
    d^{(1)} = & - b^{(1)} ~, \nonumber \\
    d^{(2)} = & - b^{(2)} - \frac{8}{9} ~, \nonumber \\
    d^{(3)} = & - b^{(3)} - \frac{527}{48} + \frac{8}{9}\zeta_2 - \frac{1}{9}\zeta_3 + \frac{4}{9}\zeta_2\ln(2) + \left(\frac{137}{216} - \frac{2}{9}\zeta_2\right)N_L \nonumber \\
    & + \left(\frac{65}{216} + \frac{4}{9}\zeta_2\right)N_H ~.
\end{align}
In addition, we obtain the relation between the scale-invariant mass and the $\overline{\mathrm{MS}}$ mass at an arbitrary renormalisation scale $\mu$ as
\begin{equation}
\label{eq:msbar_mass_running}
    \frac{m^{\overline{\mathrm{MS}}}(\mu)}{m^\mathrm{inv}} = 1 + f^{(1)}(\mu) a(\mu) + f^{(2)}(\mu) a^2(\mu) + f^{(3)}(\mu) a^3(\mu) + \mathcal{O}\left(a^4(\mu)\right)~,
\end{equation}
with the coefficients
\begin{align}
     f^{(1)}(\mu) = & - \ln_\mu ~, \nonumber \\
     f^{(2)}(\mu) = & \left(-\frac{101}{24} + \frac{5}{36} N_F\right) \ln_\mu + \left(-\frac{7}{8} + \frac{1}{12} N_F\right)\ln_\mu^2 ~, \nonumber \\
     f^{(3)}(\mu) = & \left(-\frac{1249}{64} + \left(\frac{277}{216} + \frac{5}{6}\zeta_3 \right) N_F + \frac{35}{1296} N_F^2 \right) \ln_\mu \nonumber \\ 
     & + \left(-\frac{1013}{96} + \frac{193}{144} N_F - \frac{5}{216} N_F^2 \right) \ln_\mu^2 + \left(-\frac{21}{16} + \frac{2}{9} N_F - \frac{1}{108} N_F^2\right) \ln_\mu^3
\end{align}
and the abbreviation $\ln_\mu = \ln\left(\frac{\mu^2}{(m^\mathrm{inv})^2}\right)$, whereas the ratio between the on-shell mass and the $\overline{\mathrm{MS}}$ mass at an arbitrary renormalisation scale $\mu$ is given by
\begin{equation}
\label{eq:OSmass_over_MSmass_series_expansion}
    \frac{m^\mathrm{OS}}{m^{\overline{\mathrm{MS}}}(\mu)} = 1 + h^{(1)}(\mu) a(\mu) + h^{(2)}(\mu) a^2(\mu) + h^{(3)}(\mu) a^3(\mu) + \mathcal{O}\left(a^4(\mu)\right)
\end{equation}
and the coefficients
\begin{align}
    h^{(1)} = & - b^{(1)} + \ln_{\mu m}\nonumber \\
    h^{(2)} = & - b^{(2)} - \frac{8}{9} + \left(\frac{173}{24} - \frac{13}{36}N_F\right)\ln_{\mu m} + \left(\frac{15}{8} - \frac{1}{12}N_F\right)\ln_{\mu m}^2\nonumber \\
    h^{(3)} = & -\frac{527}{48} + \frac{8}{9}\zeta_2 - \frac{1}{9}\zeta_3 + \frac{4}{9}\zeta_2\ln(2) + \left(\frac{137}{216} - \frac{2}{9}\zeta_2\right)N_L + \left(\frac{65}{216} + \frac{4}{9}\zeta_2\right)N_H \nonumber \\
    & + \left [\frac{22927}{288} + \frac{26}{3}\zeta_2 - \frac{13}{12}\zeta_3 + \frac{13}{3}\zeta_2\ln(2) + \left(-\frac{1189}{144} - \frac{47}{18}\zeta_2 - \frac{7}{9}\zeta_3 - \frac{2}{9}\zeta_2\ln(2)\right)N_L \right. \nonumber \\
    & + \left(-\frac{1657}{144} + \frac{35}{9}\zeta_2 - \frac{7}{9}\zeta_3 - \frac{2}{9}\zeta_2\ln(2)\right)N_H + \left(\frac{143}{324}-\frac{1}{9}\zeta_2\right)N_L N_H + \nonumber \\
    & + \left. \left(\frac{89}{648} + \frac{1}{9}\zeta_2\right)N_L^2 + \left(\frac{197}{648} - \frac{2}{9}\zeta_2\right)N_H^2\right]\ln_{\mu m} \nonumber \\
    & + \left[\frac{799}{32} - \frac{385}{144}N_L - \frac{385}{144}N_H + \frac{13}{108}N_L N_H + \frac{13}{216}N_L^2 + \frac{13}{216}N_H^2 \right] \ln^2_{\mu m} \nonumber \\
    & + \left[\frac{65}{16} - \frac{7}{18}N_L - \frac{7}{18}N_H  + \frac{1}{54} N_L N_H + \frac{1}{108} N_L^2 + \frac{1}{108} N_H^2\right] \ln^3_{\mu m}
\end{align}
with the abbreviation $\ln_{\mu m} = \ln\left(\frac{\mu^2}{(m^{\overline{\mathrm{MS}}}(\mu))^2}\right)$.\newline
In the calculation of differential cross-sections, Eqs.~\eqref{eq:scale_invariant_mass_series_expansion} and \eqref{eq:on_shell_mass_series_expansion} are used to convert between the on-shell mass and the scale-invariant mass, so that both masses can be used as input parameters. Eq.~\eqref{eq:msbar_mass_running} is used to calculate the $\msbar$ mass using the scale-invariant mass.\newline

The matrix elements in the on-shell scheme, $\mathcal{M}^\mathrm{OS}$, and in the $\overline{\mathrm{MS}}$ scheme $,\mathcal{M}^{\overline{\mathrm{MS}}}$, if they were calculated to all orders, must satisfy
\begin{equation}
    \mathcal{M}^\mathrm{OS}(m^\mathrm{OS}) = \mathcal{M}^{\overline{\mathrm{MS}}}(m^{\overline{\mathrm{MS}}}(\mu)) ~.
\end{equation} 
Both matrix elements can be written as a series expansion in terms of the coupling $a$
\begin{align}
\label{eq:OSmatrixelement_series_expansion}
    \mathcal{M}^\mathrm{OS}(m^\mathrm{OS}) = & \sum_{n=1}^\infty \mathcal{M}^{\mathrm{OS},(n)}(m^\mathrm{OS}) a^n(\mu) ~, \\
\label{eq:MSmatrixelement_series_expansion}
    \mathcal{M}^{\overline{\mathrm{MS}}}(m^{\overline{\mathrm{MS}}}(\mu)) = & \sum_{n=1}^\infty \mathcal{M}^{\overline{\mathrm{MS}},(n)}(m^{\overline{\mathrm{MS}}}(\mu)) a^n(\mu) ~.
\end{align}
Combining these equations with the relation between the on-shell mass $m^\mathrm{OS}$ and the $\overline{\mathrm{MS}}$ mass $m^{\overline{\mathrm{MS}}}(\mu)$ in Eq.~\eqref{eq:OSmass_over_MSmass_series_expansion} yields 
\begin{align}
    & \mathcal{M}^{\overline{\mathrm{MS}}}(m^{\overline{\mathrm{MS}}}(\mu)) = \mathcal{M}^\mathrm{OS}(m^\mathrm{OS}) \nonumber \\
    & = \sum_{n=0}^\infty \left[ \frac{1}{n!} \left. \left(\frac{\mathrm{d}}{\mathrm{d} m^\mathrm{OS}}\right)^n \mathcal{M}^\mathrm{OS}(m^\mathrm{OS}) \right|_{m^\mathrm{OS} = m^{\overline{\mathrm{MS}}}(\mu)} \left(m^{\overline{\mathrm{MS}}}(\mu) - m^\mathrm{OS} \right)^n \right] \nonumber \\
    & = \sum_{n=0}^\infty \left[ \frac{1}{n!} \left. \left(\frac{\mathrm{d}}{\mathrm{d} m^\mathrm{OS}}\right)^n \left(\mathcal{M}^{\mathrm{OS},(1)}(m^\mathrm{OS}) + \mathcal{M}^{\mathrm{OS},(2)}(m^\mathrm{OS}) + \mathcal{O}\left(a^3(\mu)\right) \right) \right|_{m^\mathrm{OS} = m^{\overline{\mathrm{MS}}}(\mu)} \right. \nonumber \\
    & \qquad \qquad\times \left. \left(m^{\overline{\mathrm{MS}}}(\mu)h^{(1)}(\mu) a(\mu) + m^{\overline{\mathrm{MS}}}(\mu)h^{(2)}(\mu)a^2(\mu) + \mathcal{O}\left(a^3(\mu)\right)\right)^n \right] \nonumber \\
    & = \mathcal{M}^{\mathrm{OS},(1)}(m^{\overline{\mathrm{MS}}}(\mu)) + \mathcal{M}^{\mathrm{OS},(2)}(m^{\overline{\mathrm{MS}}}(\mu)) \nonumber \\
    & \qquad + m^{\overline{\mathrm{MS}}}(\mu)h^{(1)}(\mu)a(\mu) \left. \frac{\mathrm{d} \mathcal{M}^{\mathrm{OS},(1)}(m^\mathrm{OS})}{\mathrm{d} m^\mathrm{OS}} \right|_{m^\mathrm{OS} = m^{\overline{\mathrm{MS}}}(\mu)} + \mathcal{O}\left(a^3(\mu)\right) ~,
\end{align}
which results in the first two coefficients of the matrix element in the $\overline{\mathrm{MS}}$ scheme being
\begin{align}
    \mathcal{M}^{\overline{\mathrm{MS}},(1)} (m^{\overline{\mathrm{MS}}}(\mu)) = & \mathcal{M}^{\mathrm{OS},(1)}(m^{\overline{\mathrm{MS}}}(\mu)) ~, \nonumber \\
    \mathcal{M}^{\overline{\mathrm{MS}},(2)} (m^{\overline{\mathrm{MS}}}(\mu)) = & \mathcal{M}^{\mathrm{OS},(2)}(m^{\overline{\mathrm{MS}}}(\mu)) +  m^{\overline{\mathrm{MS}}}(\mu)h^{(1)}(\mu)a(\mu) \left. \frac{\mathrm{d} \mathcal{M}^{\mathrm{OS},(1)}(m^\mathrm{OS})}{\mathrm{d} m^\mathrm{OS}} \right|_{m^\mathrm{OS} = m^{\overline{\mathrm{MS}}}(\mu)} ~.
\end{align}
Therefore, when calculating the leading-order matrix element in the $\overline{\mathrm{MS}}$ scheme, it is sufficient to use the leading-order matrix element from the on-shell scheme, replacing the on-shell mass with the $\overline{\mathrm{MS}}$ mass.

\subsubsection{Flavour schemes}
\label{sec:flavour_schemes}
Since we are interested in the effects of massive bottom quarks on the differential leading-order cross section of Higgs boson pair production via gluon fusion, we use the four-flavour scheme (4FS). In this scheme both the top and bottom quarks are excluded from the parton distribution functions and are treated as massive particles, which allows for the consistent use of top and bottom quark masses in the matrix element. 
Moreover, our goal is to maintain consistency with the renormalisation group evolution of the strong coupling constant and its value at the Z-boson mass $\alpha_s(M_Z)$ used in the fit of the parton distribution functions \texttt{PDF4LHC21\_40\_nf4}. To do this, we make use of \texttt{LHAPDF6} to obtain values of $\alpha_s(\mu)$ for different renormalisation scales $\mu$ from the parton distribution functions, ensuring that the same number of active quark flavours and the same order of expansion for the QCD $\beta$-function are used throughout the calculation. 
In addition, we use Eqs.~\eqref{eq:scale_invariant_mass_series_expansion} and \eqref{eq:on_shell_mass_series_expansion} to convert between the on-shell and scale-invariant masses for both the top and the bottom quark, where we set the number of light active quark flavours to $N_L=4$ and the number of heavy active quark flavours to $N_H=0$. In Eq.~\eqref{eq:msbar_mass_running} for the renormalisation group evolution of the $\overline{\mathrm{MS}}$ mass, we use $N_F=4$ active quark flavours to remain consistent with the four-flavour scheme.\newline

On the other hand, the characteristic energy scale of the scattering process, $m_{HH}/2$, is well above the bottom quark pair production threshold of $2m_b$, making it reasonable to describe the bottom quark as an active quark flavour. Such a description is provided by the five-flavour scheme (5FS), in which the bottom quark is part of the parton distribution functions and not decoupled from the renormalisation group evolution of the strong coupling constant. Unfortunately, the 5FS assumes that the bottom quark is massless, preventing us from studying the effects caused by the mass of the bottom quark. One method of circumventing this problem is by introducing two different bottom quarks as done in Ref.~\cite{Czakon:2024ywb}. Here, one of the bottom quarks is massless, part of the PDFs and contributes to the renormalisation group evolution of the strong coupling, while the bottom quark that appears in the closed fermion loops is massive, does not appear in the PDFs and is decoupled from the renormalisation group evolution of the strong coupling.

Therefore, we have performed a second computation using the 5FS and the ``mixed" treatment of the bottom quark mass described above. In an effort to be consistent with the values of the strong coupling and the number of active quark flavours used in the fit of the parton distribution functions \texttt{PDF4LHC21\_40},  we again take the strong coupling $\alpha_s(\mu)$ from the PDFs and use $N_L=5$, $N_H=0$ and $N_F=5$ in Eqs.~\eqref{eq:scale_invariant_mass_series_expansion}, \eqref{eq:on_shell_mass_series_expansion} and \eqref{eq:msbar_mass_running}.

\subsection{Results}
In this section, we present our results for the effects of massive bottom quark loops on the leading order cross section for Higgs boson pair production via gluon fusion in proton-proton scattering. The beam energy is set to $7\,\mathrm{TeV}$ per proton beam. For the quark masses, we use an on-shell mass of $m_t^\mathrm{OS} = 173\,\mathrm{GeV}$ for the top quark and a scale invariant mass of $m_b^\mathrm{inv} = 4.18\,\mathrm{GeV}$ for the bottom quark as input parameters for the calculations. Using Eqs.~\eqref{eq:scale_invariant_mass_series_expansion}, \eqref{eq:on_shell_mass_series_expansion} and \eqref{eq:msbar_mass_running} these input parameters allow us to compute the on-shell and $\overline{\mathrm{MS}}$ masses for both the top and bottom quark.

If not otherwise specified, we use a 4FS as characterized in Section~\ref{sec:flavour_schemes} to deal with the massive bottom quarks. Moreover, we set the factorisation scale $\facscale$ and the renormalisation scale $\renscale$ to the dynamic scale $m_{HH}/2$, which is also used for the mass renormalisation scales $\mu_{m_t}$ and $\mu_{m_b}$ of the $\overline{\mathrm{MS}}$ masses $m_t^{\overline{\mathrm{MS}}}(\mu_{m_t})$ and $m_b^{\overline{\mathrm{MS}}}(\mu_{m_b})$.

\subsubsection{Effect of massive bottom quarks}
\label{sec:effects_of_massive_bottom_quarks}
\begin{figure}[t]
    \centering
    \includegraphics[width=1.0\linewidth]{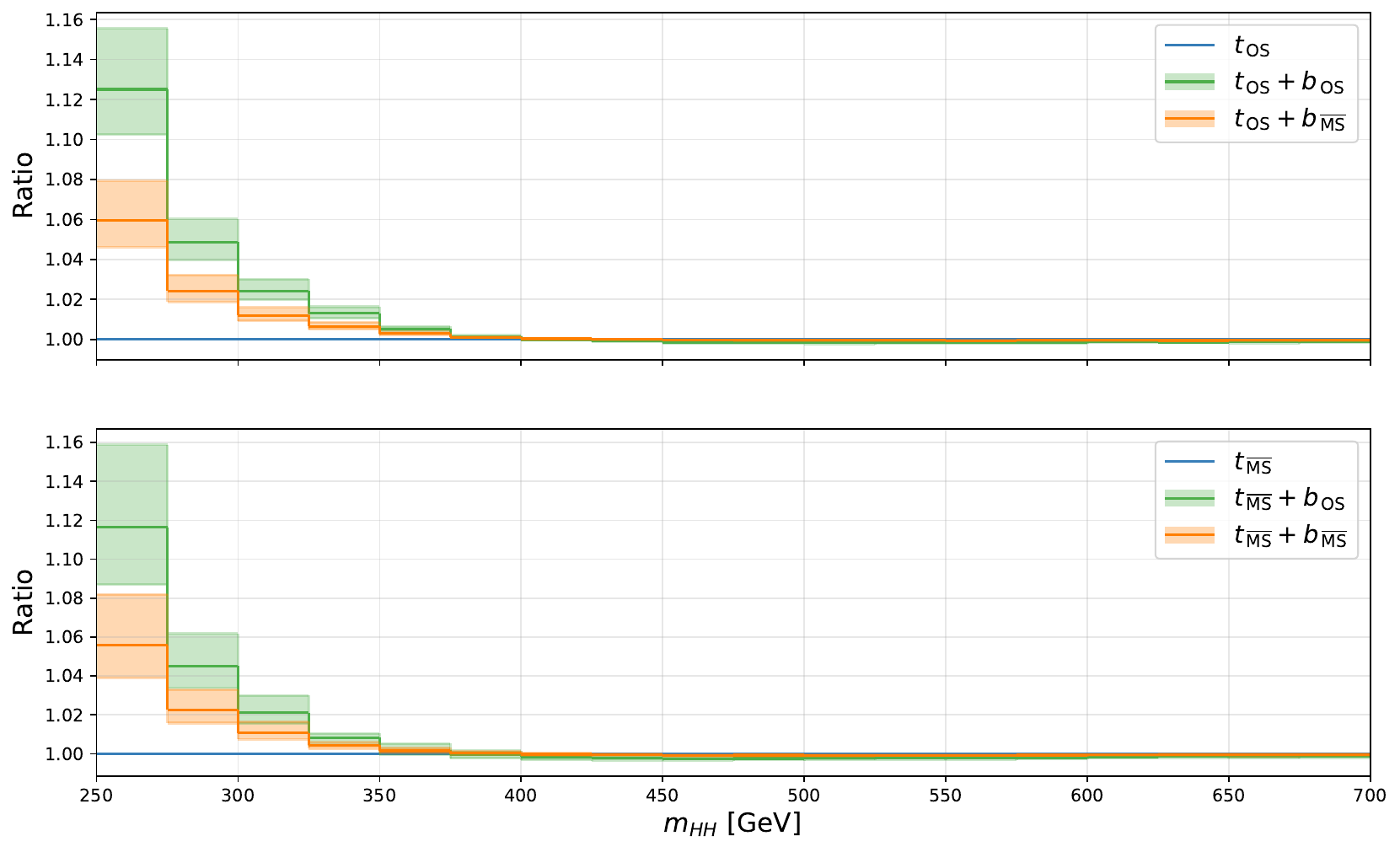}
    \caption{Effects of massive bottom quarks on the Higgs boson pair invariant mass distribution, $\frac{\mathrm{d}\sigma(pp \rightarrow HH)}{\mathrm{d}m_{HH}}$. The upper plot shows the ratio between the differential cross section computed without a massive bottom quark (blue baseline), and with a massive bottom quark renormalised in the on-shell scheme (green) or $\overline{\mathrm{MS}}$ scheme (orange), whereby the top quark mass is renormalised in the on-shell scheme. The lower plot shows the same ratios, with the top quark mass being renormalised in the $\overline{\mathrm{MS}}$ scheme. The scale uncertainties due to the massive bottom quark were calculated using Eq.~\eqref{eq:b-scale-uncertainties}.}
    \label{fig:bmasseffect}
\end{figure}

The impact of massive bottom quarks on the Higgs boson pair invariant mass distribution,  $\frac{\mathrm{d}\sigma(pp \rightarrow HH)}{\mathrm{d}m_{HH}}$, is depicted in Figure~\ref{fig:bmasseffect}, which shows the ratios of differential cross sections computed once with and once without massive bottom quark loops.
More precisely, the upper plot in Figure~\ref{fig:bmasseffect} shows this ratio for the top quark mass being renormalised in the on-shell scheme and the bottom quark mass being renormalised once in the on-shell and once in the $\overline{\mathrm{MS}}$ scheme, while the lower plot in Figure~\ref{fig:bmasseffect} shows these ratios for the top quark mass being renormalised in the $\overline{\mathrm{MS}}$ scheme.

In both cases, regardless of whether the top quark mass is renormalised using the on-shell or the $\overline{\mathrm{MS}}$ scheme, we obtain similar results: When the bottom quark mass is renormalised in the $\overline{\mathrm{MS}}$ scheme and a bin size of $25\,\mathrm{GeV}$ is used, the differential cross section near the Higgs pair production threshold increases by approximately $6\,\%$, taking massive bottom quark loops into account. If, on the other hand, the bottom quark mass is renormalised in the on-shell scheme, the differential cross section near the Higgs pair production threshold increases by approximately $12\,\%$.\newline

The uncertainties shown in Figure~\ref{fig:bmasseffect} represent the scale uncertainties associated with the bottom quark. To compute these scale uncertainties, we use a 3-point scale variation for each mass renormalised in the $\overline{\mathrm{MS}}$ scheme and a 7-point scale variation for the factorisation and renormalisation scales; i.\,e.\ we compute the differential cross section for all
\begin{equation*}
    \mu_{m_t} \in \frac{m_{HH}}{2}\{0.5,1,2\} \qquad \mathrm{and} \qquad \mu_{m_b} \in \frac{m_{HH}}{2}\{0.5,1,2\} ~,
\end{equation*}
if the respective mass is renormalised in the $\overline{\mathrm{MS}}$ scheme, as well as for all pairs of
\begin{equation}
    (\facscale,\renscale) \in \frac{m_{HH}}{2}\{(0.5,0.5),(0.5,1),(1,0.5),(1,1),(1,2),(2,1),(2,2)\}
    \label{eq:7-point}
\end{equation}
and then construct the resulting envelope.

In addition, we split the total differential cross section into a contribution due to the top quark, an interference term and a contribution due to the bottom quark
\begin{equation}
    \frac{\mathrm{d}\sigma^\mathrm{total}}{\mathrm{d}m_{HH}} = \frac{\mathrm{d}\sigma^t}{\mathrm{d}m_{HH}} + \frac{\mathrm{d}\sigma^\mathrm{int}}{\mathrm{d}m_{HH}} + \frac{\mathrm{d}\sigma^b}{\mathrm{d}m_{HH}}
\end{equation}
and apply the scale variation only to the terms associated with the bottom quark mass
\begin{align}
    \left. \frac{\mathrm{d}\sigma^\mathrm{total}}{\mathrm{d}m_{HH}} \right|_\mathrm{bottom~scale~uncertainties} = \frac{\mathrm{d}\sigma^t}{\mathrm{d}m_{HH}} + \left[ \frac{\mathrm{d}\sigma^\mathrm{int}}{\mathrm{d}m_{HH}}  + \frac{\mathrm{d}\sigma^b}{\mathrm{d}m_{HH}} \right]_\mathrm{scale~variation} ~.
\end{align}
In this way, we can specify only the scale uncertainties attributable to the massive bottom quark, and avoid these scale uncertainties being masked by scale uncertainties associated with the contribution of the top quark, which is already known to higher orders~\cite{Baglio:2018lrj,Bagnaschi:2023rbx,Jaskiewicz:2024xkd,Davies:2025ghl,Davies:2025qjr,Chen:2026zmi}. Further, a more practical form can be obtained by rearranging the formula in the following way
\begin{align}
    \left. \frac{\mathrm{d}\sigma^\mathrm{total}}{\mathrm{d}m_{HH}} \right|_\mathrm{bottom~scale~uncertainties} = &  \left[ \frac{\mathrm{d}\sigma^t}{\mathrm{d}m_{HH}} + \frac{\mathrm{d}\sigma^\mathrm{int}}{\mathrm{d}m_{HH}}  + \frac{\mathrm{d}\sigma^b}{\mathrm{d}m_{HH}} - \frac{\mathrm{d}\sigma^t}{\mathrm{d}m_{HH}} \right]_\mathrm{scale~variation} + \frac{\mathrm{d}\sigma^t}{\mathrm{d}m_{HH}} \nonumber \\
    = & \left[ \frac{\mathrm{d}\sigma^\mathrm{total}}{\mathrm{d}m_{HH}} - \frac{\mathrm{d}\sigma^t}{\mathrm{d}m_{HH}}\right]_\mathrm{scale~variation} + \frac{\mathrm{d}\sigma^t}{\mathrm{d}m_{HH}} ~.
\label{eq:b-scale-uncertainties}
\end{align}

\begin{figure}[h!]
    \centering
    \includegraphics[width=1.0\linewidth]{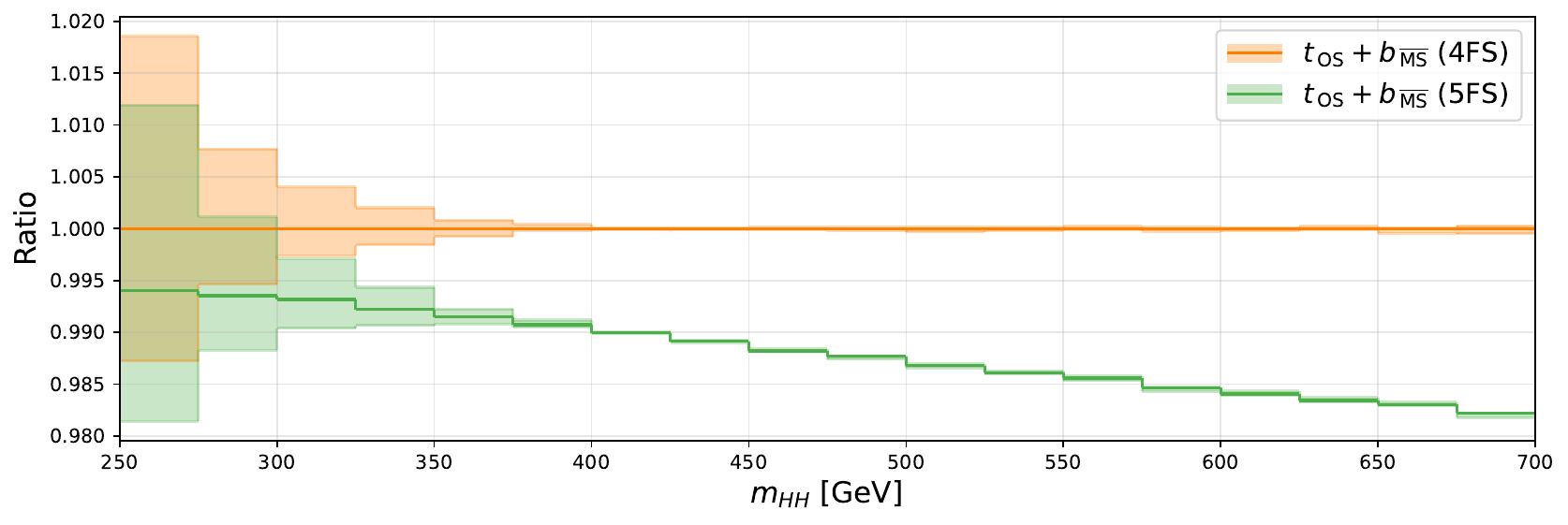}
    \caption{Difference between the 4FS and 5FS in the  $\mhh$ distribution. The orange baseline represents the differential cross section calculated in the 4FS, while the green line represents the differential cross section calculated in the 5FS. The top quark mass is renormalised in the on-shell scheme and the bottom quark mass is renormalised in the $\overline{\mathrm{MS}}$ scheme. The scale uncertainties due to the massive bottom quark were calculated as given in Eq.~\eqref{eq:b-scale-uncertainties}.}
    \label{fig:ratio_FSs}

    \vspace{1cm} 

    \centering
    \includegraphics[width=1.0\linewidth]{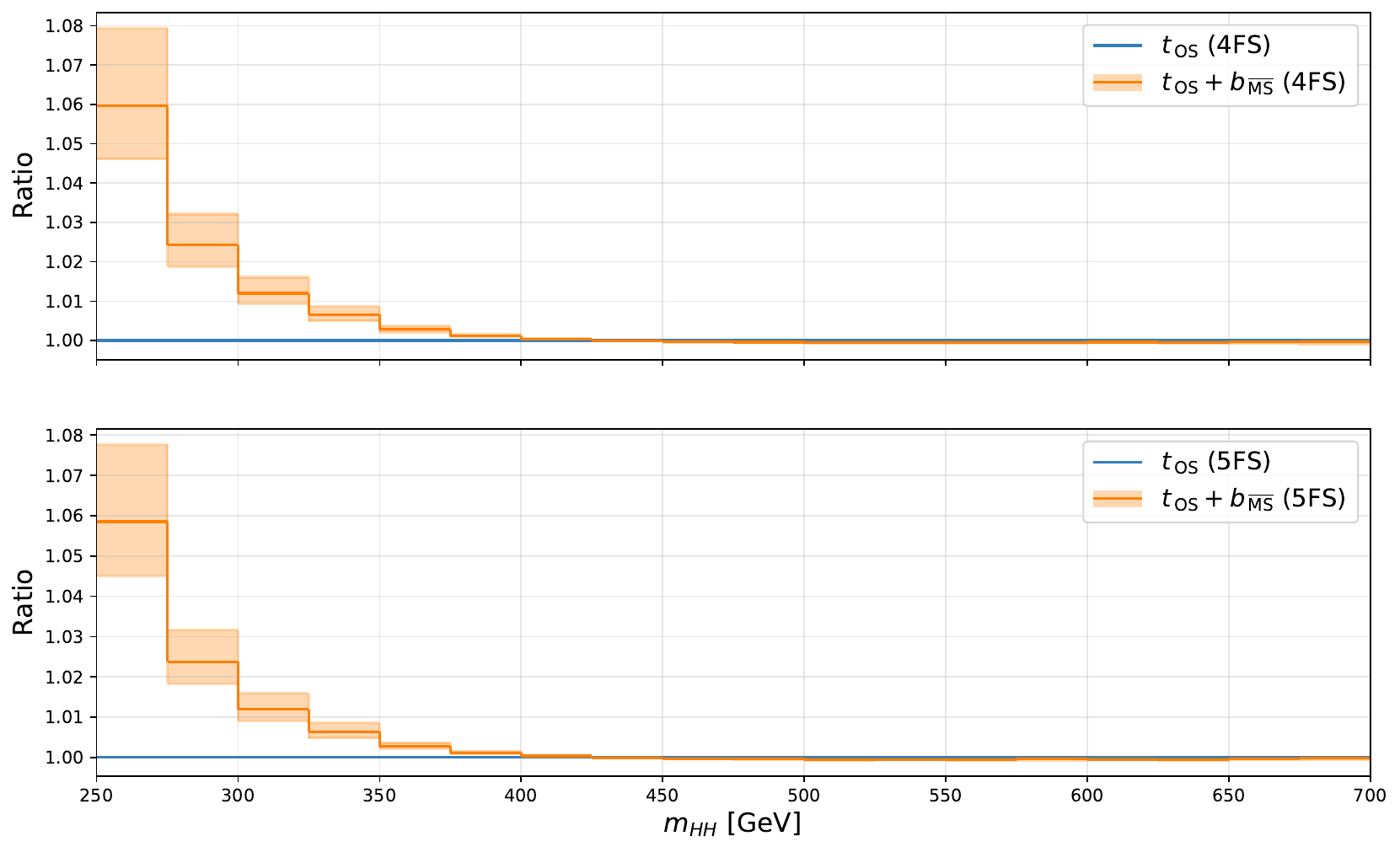}
    \caption{Influence of different flavour number schemes on the effect of massive bottom quarks on the $\mhh$ distribution. The upper plot shows the ratio between the differential cross sections calculated in the 4FS once with (orange) and once without massive bottom quarks (blue baseline). The lower plot shows the same ratios calculated using the 5FS. The top quark mass is renormalised in the on-shell scheme and the bottom quark mass is renormalised in the $\overline{\mathrm{MS}}$ scheme. The scale uncertainties due to the massive bottom quark were calculated using Eq.~\eqref{eq:b-scale-uncertainties}.}
    \label{fig:bmasseffects_FSs}
\end{figure}

\subsubsection{Comparison of the four- and the five-flavour scheme}
The results presented in Section~\ref{sec:effects_of_massive_bottom_quarks} were calculated using the 4FS. To investigate the effect of the flavour number scheme, we repeated the calculations for the 5FS described in Section~\ref{sec:flavour_schemes}, obtaining the results presented in Figures~\ref{fig:ratio_FSs} and \ref{fig:bmasseffects_FSs}.\newline

Figure~\ref{fig:ratio_FSs} shows the ratio between the differential cross sections computed in the 4FS and 5FS, with the top quark mass being renormalised in the on-shell scheme and the bottom quark mass being renormalised in the $\overline{\mathrm{MS}}$ scheme. We see that the difference between both flavour number schemes is less than $1\,\%$ near the Higgs pair production threshold but increases with increasing Higgs pair invariant mass $m_{HH}$.

Figure~\ref{fig:bmasseffects_FSs}, on the other hand, shows the ratio between the differential cross sections calculated once with and once without massive bottom quarks, where the top quark mass is renormalised in the on-shell scheme and the bottom quark mass is renormalised in the $\overline{\mathrm{MS}}$ scheme. The upper plot in Figure~\ref{fig:bmasseffects_FSs} shows this ratio, calculated using the 4FS, while the lower plot shows the same ratio, calculated using the 5FS. As can be seen, the effect of the massive bottom quark is almost identical in both flavour number schemes, showing that the flavour number scheme used is not decisive for the effects of massive bottom quarks investigated here.

\subsubsection{Effect of anomalous trilinear Higgs couplings}
To investigate the influence of the trilinear Higgs coupling on the effect of massive bottom quarks on the invariant mass distribution of Higgs boson pairs, we make use of the electroweak chiral Lagrangian (EWChL), also called Higgs Effective Field Theory (HEFT)~\cite{Alonso:2012px, Buchalla:2013rka,Buchalla:2018yce}. The term of interest within this Lagrangian is given by
\begin{equation}
    \mathcal{L}_\mathrm{EWChL} \supset - c_{HHH} \frac{m_H^2}{v^2} H^3 ~,
\end{equation}
where $c_{HHH}$, $m_H$, $v$, and $H$ denote the anomalous trilinear Higgs coupling, the mass of the Higgs boson, the vacuum expectation value, and the Higgs boson field, respectively.\newline

\begin{figure}
    \centering
    \includegraphics[width=1.0\linewidth]{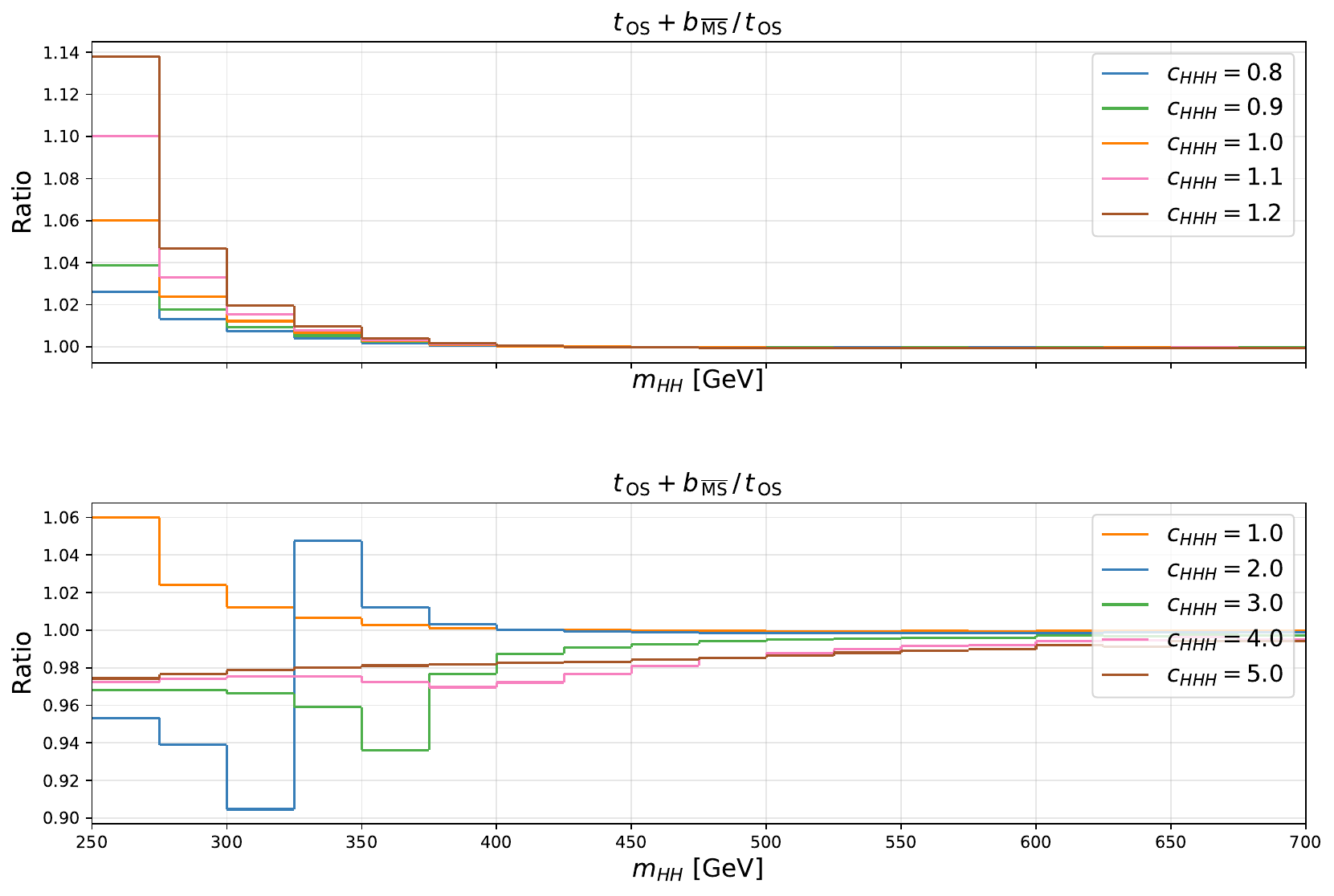}
    \caption{Effect of massive bottom quarks on the invariant mass distribution of Higgs boson pairs for different values of the anomalous trilinear Higgs coupling $c_{HHH}$. The upper plot shows the ratio between the differential cross section computed once with only the top quark mass renormalised in the OS scheme and once with the top quark renormalised in the OS scheme and the bottom quark mass renormalised in the $\overline{\mathrm{MS}}$ scheme, for different values of $c_{HHH}$ around its SM value of $c_{HHH} = 1$. The lower plot shows the same ratio for larger values of the anomalous trilinear Higgs coupling.}
    \label{fig:bmasseffect_chhh}
\end{figure}

Figure~\ref{fig:bmasseffect_chhh} shows how the effect of massive bottom quarks on the invariant mass distribution of Higgs boson pairs varies for different values of the anomalous trilinear Higgs coupling when the top quark mass is renormalised in the OS scheme and the bottom quark mass in the $\overline{\mathrm{MS}}$ scheme.

The upper plot of Figure~\ref{fig:bmasseffect_chhh} shows the variation of the anomalous trilinear Higgs coupling around its Standard Model value, $c_{HHH} = 1$. We observe that the effect of massive bottom quarks on the invariant mass distribution near the Higgs boson pair production threshold increases significantly as $c_{HHH}$ increases for values of the trilinear coupling in the vicinity of $c_{HHH}=1$.

A different pattern emerges in the lower plot of Figure~\ref{fig:bmasseffect_chhh}, which illustrates the influence of massive bottom
quarks at larger values of the anomalous trilinear Higgs coupling. Here we see that, as the trilinear Higgs coupling increases, the effect of the massive bottom quarks in decreases the cross section near the Higgs boson pair production threshold compared to the SM. Near the point of maximal destructive interference between box- and triangle diagrams, $c_{HHH} \simeq 2$, it shows a pronounced dip-peak structure near the top quark  pair production threshold.

\subsubsection{Effect of different scale variations on scale uncertainties}
To calculate scale uncertainties, the factorisation scale $\facscale$ and the renormalisation scale $\renscale$ are varied around their central values. In our case, we used half the Higgs pair invariant mass $m_{HH}/2$ as central value for both scales and varied the scales in a seven-point scale variation as listed in Eq.~\eqref{eq:7-point}.

In addition, for every mass $m^{\overline{\mathrm{MS}}}$ renormalised in the $\overline{\mathrm{MS}}$ scheme, an additional mass renormalisation scale $\mu_m$ arises, which can also be varied. Thus, in the previous calculations, we varied each mass renormalisation scale $\mu_m$ independently along a three-point scale variation around half the Higgs pair invariant mass $m_{HH}/2$, the same central value as for the factorisation and the renormalisation scale.\newline

This approach has the drawback that a large number of variations must be calculated if each mass renormalisation scale is varied independently. For example, 21 variations were calculated when one mass was renormalised in the $\overline{\mathrm{MS}}$ scheme to determine the scale uncertainties; while 63 variations were calculated when both quark masses were renormalised in the $\overline{\mathrm{MS}}$ scheme. If, on the other hand, the mass renormalisation scales are not varied independently, this problem does not arise. We have therefore investigated how the scale uncertainties behave in this case.

\begin{figure}
    \centering
    \includegraphics[width=1.0\linewidth]{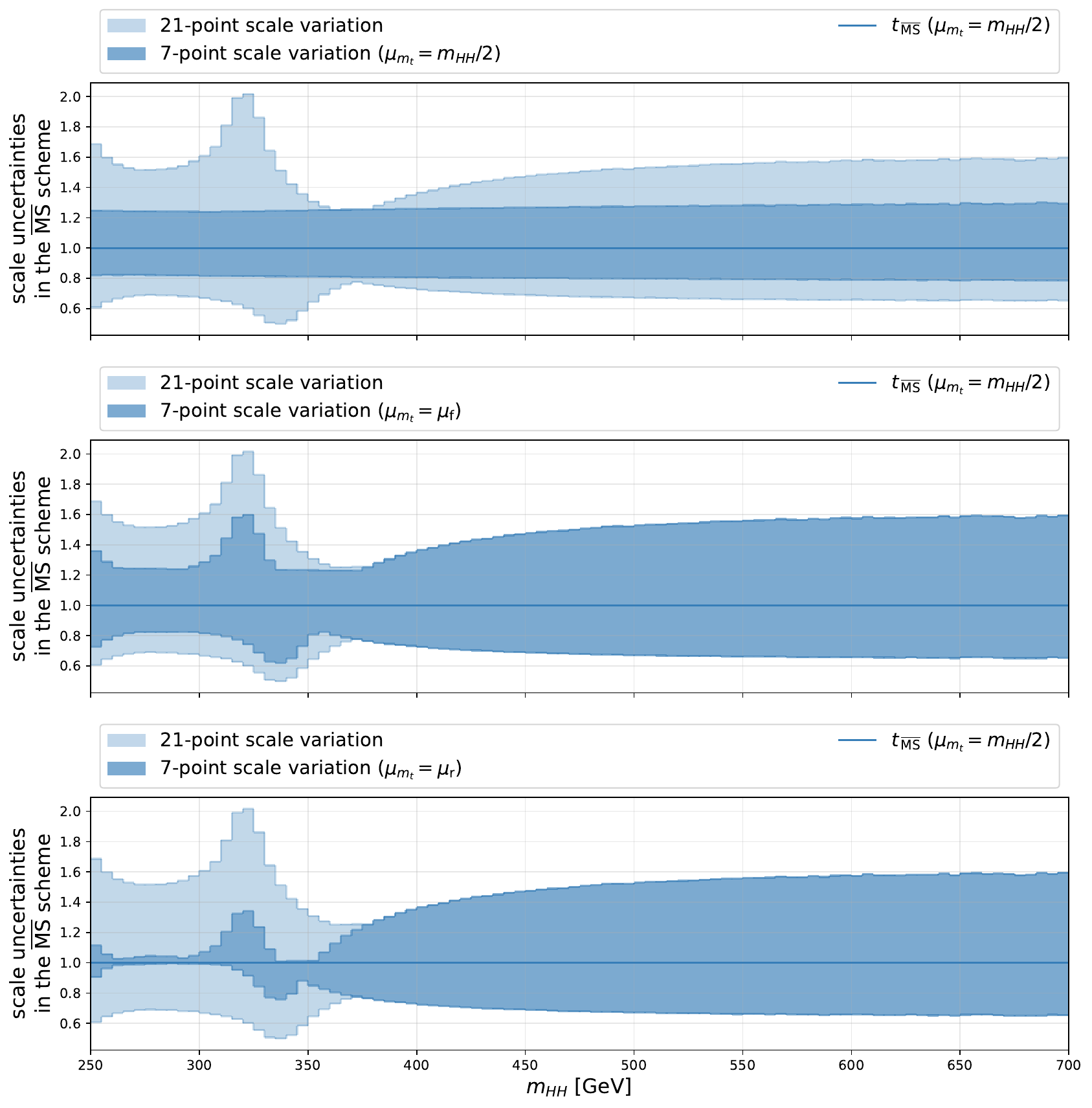}
    \caption{Different scale uncertainties of the $\mhh$ distribution. 
    Each plot shows the same scale uncertainties for a 21-point variation of the differential cross-section (light blue), which was determined by renormalising the top quark mass according to the $\overline{\mathrm{MS}}$ scheme, whilst neglecting contributions from massive bottom quarks. In addition, the three plots show the scale uncertainties obtained from the 7-point scale variation of the factorisation and renormalisation scales $\facscale$ and $\renscale$ (dark blue), with the top mass renormalisation scale $\mu_{m_t}$ set to $m_{HH}/2$, $\facscale$ and $\renscale$ respectively. All scale uncertainties are normalised to the baseline obtained by renormalising the top quark mass in the $\msbar$ scheme with $\mu_{m_t}=\mhh/2$.}
    \label{fig:scale_uncertainties_only_top}
\end{figure}

Figure~\ref{fig:scale_uncertainties_only_top} shows the scale uncertainties of the $\mhh$ differential cross section, where the top quark mass is renormalised in the $\overline{\mathrm{MS}}$ scheme and contributions from massive bottom quarks are neglected. All three plots in Figure~\ref{fig:scale_uncertainties_only_top} show the same scale uncertainties, which result from a 21-point scale variation, i.\,e.\, in which the factorisation and renormalisation scales are varied in a 7-point scale variation, whilst the renormalisation scale of the top mass is varied independently in a 3-point scale variation.  Furthermore, each plot of Figure~\ref{fig:scale_uncertainties_only_top} shows an additional scale uncertainty resulting from a 7-point scale variation in which the top mass renormalisation scale is not varied independently: The upper plot also shows the scale uncertainties arising from the seven-point scale variation of the renormalisation and factorisation scales, when the renormalisation scale of the top quark mass is set to half the invariant mass of the Higgs pair, $\mu_{m_t} = m_{HH}/2$. In contrast, the central and lower plots show the scale uncertainties arising from a seven-point scale variation of the factorisation and renormalisation scale, in which the top mass renormalisation scale is set equal to the factorisation scale $\mu_{m_t} = \facscale$ or the renormalisation scale $\mu_{m_t} = \renscale$, respectively.

Looking at the upper plot in Figure~\ref{fig:scale_uncertainties_only_top}, it can be seen that if the top mass renormalisation scale is not varied at all, i.\,e.\ $\mu_{m_t} = m_{HH}/2$, the scale uncertainty is approximately constant, and apart from a small region slightly above twice the on-shell top quark mass $2m_t^\mathrm{OS}$ the scale uncertainty is significantly lower than in the 21-point scale variation. The central and lower plots, on the other hand, show that the scale uncertainties of the 21-point and 7-point scale variations hardly differ for $m_{HH} \gg 2 m_t^\mathrm{OS}$ when the top mass renormalisation scale is varied together with the factorisation or the renormalisation scale. For $m_{HH} \lesssim 2 m_t^\mathrm{OS}$, however, the scale uncertainty of the 7-point scale variation is significantly lower than the scale uncertainty of the 21-point scale variation.

In particular, we would like to draw attention to the scale uncertainties for $m_{HH} \lesssim 2 m_t^\mathrm{OS}$ that arise when the renormalisation scale and the top mass renormalisation scale are varied together, i.\,e.\ $\mu_{m_t} = \renscale$: In this case, the scale uncertainties appear to be artificially small, particularly as they are significantly smaller than in the 7-point scale variation, in which the renormalisation scale of the top quark mass is not varied at all; see the upper plot in Figure~\ref{fig:scale_uncertainties_only_top}.\newline

A similar analysis was carried out for the scale uncertainties of the differential cross section arising from the massive bottom quark, whereby the top quark mass was renormalised in the on-shell scheme and the bottom quark mass was renormalised in the $\overline{\mathrm{MS}}$ scheme. Since, in this case, half the invariant mass of the Higgs pair was well above twice the on-shell bottom quark mass, only negligible differences were observed between the scale uncertainties resulting from the 21-point and various 7-point scale variations.

\subsubsection{Influence of the order of $\alpha_s$ in the mass conversion and the renormalisation group evolution of quark masses}
Although the cross section is only taken into account to leading order, we have included the mass conversion between the on-shell mass and the scale-invariant mass given in Eqs.~\eqref{eq:scale_invariant_mass_series_expansion} and \eqref{eq:on_shell_mass_series_expansion}, as well as the renormalisation group evolution of the $\overline{\mathrm{MS}}$ mass in Eq.~\eqref{eq:msbar_mass_running} up to next-to-next-to-leading order.\newline

\begin{figure}
    \centering
    \includegraphics[width=1.0\linewidth]{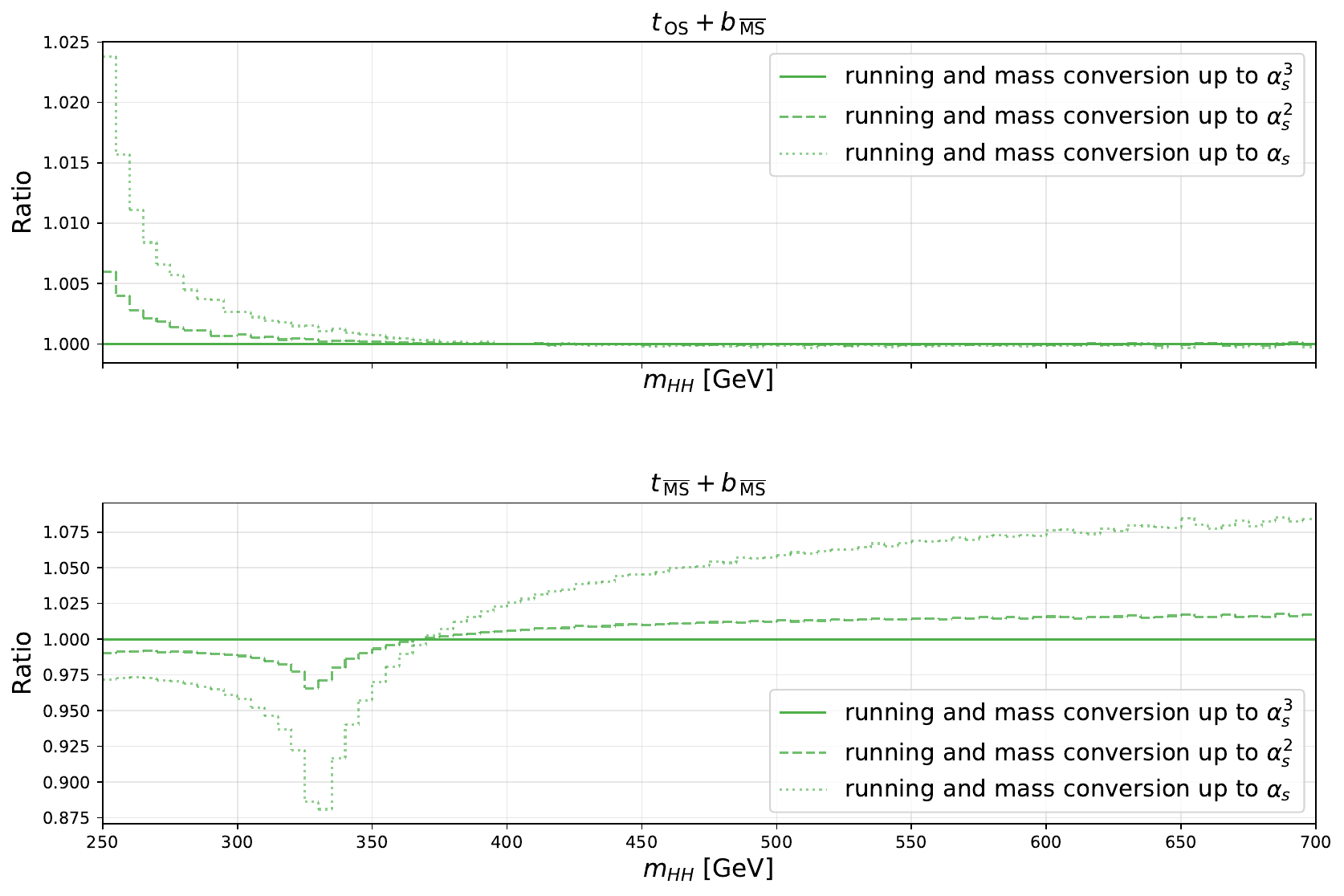}
    \caption{Effect of varying expansion orders in the conversion between the on-shell and the scale-invariant mass as well as the renormalisation group evolution of the $\overline{\mathrm{MS}}$ mass on the $\mhh$ distribution.
    The solid baseline shows the $\mhh$ distribution, in which Eqs.~\eqref{eq:scale_invariant_mass_series_expansion}, \eqref{eq:on_shell_mass_series_expansion} and \eqref{eq:msbar_mass_running} were expanded up to $\mathcal{O}\left(\alpha_s^3\right)$ to determine the masses of the top and bottom quarks, whereas the dashed and dotted lines show the same distribution in relation to this, if these equations are expanded only up to $\mathcal{O}\left(\alpha_s^2\right)$ and $\mathcal{O}\left(\alpha_s\right)$ respectively.
    The top quark mass is renormalised in the on-shell scheme in the upper plot and in the $\overline{\mathrm{MS}}$ scheme in the lower plot, whereas the bottom quark mass is renormalised in the $\msbar$ scheme in both cases.}
    \label{fig:ratio_omc}
\end{figure}

To investigate the effect of expanding to different orders in $\alpha_s$ on the differential cross-section, we have recalculated it, using Eqs.~\eqref{eq:scale_invariant_mass_series_expansion}, \eqref{eq:on_shell_mass_series_expansion} and \eqref{eq:msbar_mass_running} only up to $\mathcal{O}\left(\alpha_s\right)$ and $\mathcal{O}\left(\alpha_s^2\right)$ instead of $\mathcal{O}\left(\alpha_s^3\right)$. Accordingly, the upper plot in Figure~\ref{fig:ratio_omc} shows ratios of the differential cross section for different expansion orders, with the top quark mass renormalised in the on-shell scheme and the bottom quark mass renormalised in the $\overline{\mathrm{MS}}$ scheme. The lower plot shows the same ratios when the top quark mass is also renormalised in the $\overline{\mathrm{MS}}$ scheme.

The deviations in the low single-digit range shown in the upper plot of Figure~\ref{fig:ratio_omc} originate solely from the renormalisation group evolution of the $\overline{\mathrm{MS}}$ mass of the bottom quark, since the on-shell mass of the top quark and the scale-invariant mass of the bottom quark are used as input parameters. However, in the lower plot, the mass of the top quark is renormalised according to the $\overline{\mathrm{MS}}$ scheme. Accordingly, the on-shell mass of the top quark must be converted to the scale-invariant mass here and further evolved in accordance with the renormalisation group equation. Together with the renormalisation group evolution of the bottom quark $\overline{\mathrm{MS}}$ mass, these two effects lead to deviations in the percentage range, which are particularly large around $m_{HH}/2 \simeq m_t^\mathrm{OS}$.

\clearpage

\section{Bottom quark annihilation}
\label{sec:initial_bottom}
\subsection{Setup}
we compute the LO contribution and the NLO QCD corrections to Higgs boson pair production via bottom--antibottom annihilation, 
\begin{align}
    b\overline{b} \to H H \, .
\end{align}
Within the 5FS, the bottom quarks in the initial state and in internal propagators are treated as massless partons of the proton. The bottom quark Yukawa coupling, however, is expressed in terms of the renormalised $\overline{\mathrm{MS}}$ mass, $m_b^{\overline{\mathrm{MS}}}$. Representative diagrams at LO and NLO, including virtual corrections and real emissions, are shown in Table~\ref{tab:diagrams}.

\begin{table}[htb]
    \centering 
    \caption{Example diagrams for Higgs boson pair production via bottom-antibottom annihilation.}
    \label{tab:diagrams} 
    \begin{tabular}{>{\centering\arraybackslash}p{3cm}>{\centering\arraybackslash}p{3cm}>{\centering\arraybackslash}p{3cm}}
    \toprule
    Leading Order & \multicolumn{2}{c}{\hspace{1.5cm}Next-to-Leading Order QCD} \\
    & real emissions & virtual corrections \\
    \midrule
    \begin{tikzpicture}
    \begin{feynman}[inline = (hi)]
        \vertex (hi);
        \vertex [right = 0.6cm of hi] (ha);
        \vertex [right = 0.6cm of ha] (hb);
        \vertex [right = 0.6cm of hb] (hf);
        \vertex [above = 0.6cm of hi] (i1);
        \vertex [above = 0.6cm of hf] (f1);
        \vertex [below = 0.6cm of hi] (i2);
        \vertex [below = 0.6cm of hf] (f2);
        \diagram*{
            (i1) -- [fermion, red] (ha) -- [fermion, red] (i2), 
            (ha) -- [scalar] (hb), 
            (f1) -- [scalar] (hb) -- [scalar] (f2)
        };
    \end{feynman}
\end{tikzpicture} & 
    \begin{tikzpicture}
    \begin{feynman}[inline = (hi)]
        \vertex (hi);
        \vertex [right = 0.6cm of hi] (ha);
        \vertex [right = 0.6cm of ha] (hb);
        \vertex [right = 0.6cm of hb] (hf);
        \vertex [right = 0.3cm of hi] (hg); 
        \vertex [right = 0.3cm of ha] (hr); 
        \vertex [above = 0.6cm of hi] (i1);
        \vertex [above = 0.6cm of hf] (f1);
        \vertex [above = 0.3cm of hg] (g1); 
        \vertex [above = 0.6cm of hr] (r1); 
        \vertex [below = 0.6cm of hi] (i2);
        \vertex [below = 0.6cm of hf] (f2);
        \diagram*{
            (g1) -- [gluon] (r1),
            (i1) -- [fermion, red] (ha) -- [fermion, red] (i2), 
            (ha) -- [scalar] (hb), 
            (f1) -- [scalar] (hb) -- [scalar] (f2), 
        };
    \end{feynman}
\end{tikzpicture} 
& 
    \begin{tikzpicture}
    \begin{feynman}[inline = (hi)]
        \vertex (hi);
        \vertex [right = 0.6cm of hi] (ha);
        \vertex [right = 0.6cm of ha] (hb);
        \vertex [right = 0.6cm of hb] (hf);
        \vertex [right = 0.3cm of hi] (hg); 
        \vertex [above = 0.6cm of hi] (i1);
        \vertex [above = 0.6cm of hf] (f1);
        \vertex [above = 0.3cm of hg] (g1); 
        \vertex [below = 0.6cm of hi] (i2);
        \vertex [below = 0.6cm of hf] (f2);
        \vertex [below = 0.3cm of hg] (g2); 
        \diagram*{
            (g1) -- [gluon] (g2),
            (i1) -- [fermion, red] (ha) -- [fermion, red] (i2), 
            (ha) -- [scalar] (hb), 
            (f1) -- [scalar] (hb) -- [scalar] (f2), 
        };
    \end{feynman}
\end{tikzpicture} \\
    \bottomrule 
\end{tabular}
\end{table}

We use \texttt{POWHEG-BOX-V2}~\cite{Nason:2004rx, Frixione:2007vw, Alioli:2010xd} to compute differential distributions, with \texttt{GoSam-3}~\cite{Braun:2025afl} providing the amplitude.

As electroweak input parameters we choose the mass of the Higgs boson $m_H = 125\,\mathrm{GeV}$, the mass of the W boson $m_W = 80.36\,\mathrm{GeV}$, the mass of the Z boson $m_Z = 91.1876\,\mathrm{GeV}$, and the Fermi constant $G_F = 1.16639\times10^{-5}\,\mathrm{GeV}^{-2}$. The running bottom quark mass $m_b^{\overline{\mathrm{MS}}}(\renscale)$ at the renormalisation scale $\renscale$ is obtained from the invariant bottom quark mass $m_b^\mathrm{inv} = 4.186\,\mathrm{GeV}$ according to Eq.~(13) of Ref.~\cite{Harlander:2003ai}, with the perturbative expansion of $c(a_s)$ truncated at first order in $a_s = \alpha_s/\pi$. The strong coupling constant $\alpha_s(\mu)$ is taken from the PDF set \texttt{PDF4LHC21\_40}~\cite{PDF4LHCWorkingGroup:2022cjn}.

\subsection{Results}

\begin{table}[htb]
    \centering
    \caption{LO and NLO cross sections for bottom quark annihilation at different proton--proton centre-of-mass energies. The uncertainties correspond to renormalisation and factorisation scale uncertainties obtained from a 7-point scale variation around the dynamic central scale $\mu_\mathrm{central}=m_{HH}/2$ and the fixed central scale $\mu_\mathrm{central}=2m_H$, shown on the left and right, respectively.}
    \begin{tabular}{ccccccc}
\toprule
& \multicolumn{2}{c}{Fixed scale}
& \multicolumn{2}{c}{Running scale}\\
$\sqrt{S}$
& LO [fb] & NLO [fb]
& LO [fb] & NLO [fb] \\
\midrule
13 TeV
& $0.223^{+0.042}_{-0.046}$
& $0.217^{+0.011}_{-0.013}$
& $0.223^{+0.042}_{-0.046}$
& $0.217^{+0.011}_{-0.013}$\\[3pt]

13.6 TeV
& $0.246^{+0.047}_{-0.052}$
& $0.239^{+0.012}_{-0.015}$
& $0.246^{+0.047}_{-0.052}$
& $0.239^{+0.012}_{-0.015}$\\[3pt]

14 TeV
& $0.261^{+0.050}_{-0.055}$
& $0.255^{+0.013}_{-0.016}$
& $0.261^{+0.050}_{-0.055}$
& $0.255^{+0.013}_{-0.016}$\\
\bottomrule
\end{tabular}
    \label{tab:xsec}
\end{table}

\subsubsection{Cross section}
Table~\ref{tab:xsec} reports the LO and NLO total cross sections for three choices of the proton--proton centre-of-mass energy: $13\, \mathrm{TeV}$, $13.6\, \mathrm{TeV}$, and $14\, \mathrm{TeV}$. For each energy, we assess the scale uncertainties by performing a 7-point scale variation around both a dynamic scale choice, $\renscale = \facscale = m_{HH}/2$, where $m_{HH}$ denotes the invariant mass of the two Higgs bosons, and a fixed scale choice, $\renscale = \facscale = 2\,m_H$.

We find agreement with the values reported in Ref.~\cite{AH:2018tzg}  within $\mathcal{O}(1\%)$ by choosing the corresponding central scale, setting the invariant bottom mass to $m_b^\mathrm{inv} = 4.7\,\mathrm{GeV}$ and using the \texttt{MMHT2014} PDF set \cite{Harland-Lang:2014zoa}. The remaining discrepancy is likely caused by a different electroweak input scheme affecting both the bottom quark Yukawa coupling and the trilinear coupling of the Higgs boson.  

\subsubsection{Differential Distributions}
\begin{figure}[tb]
    \centering
    \includegraphics[width=0.48\linewidth]{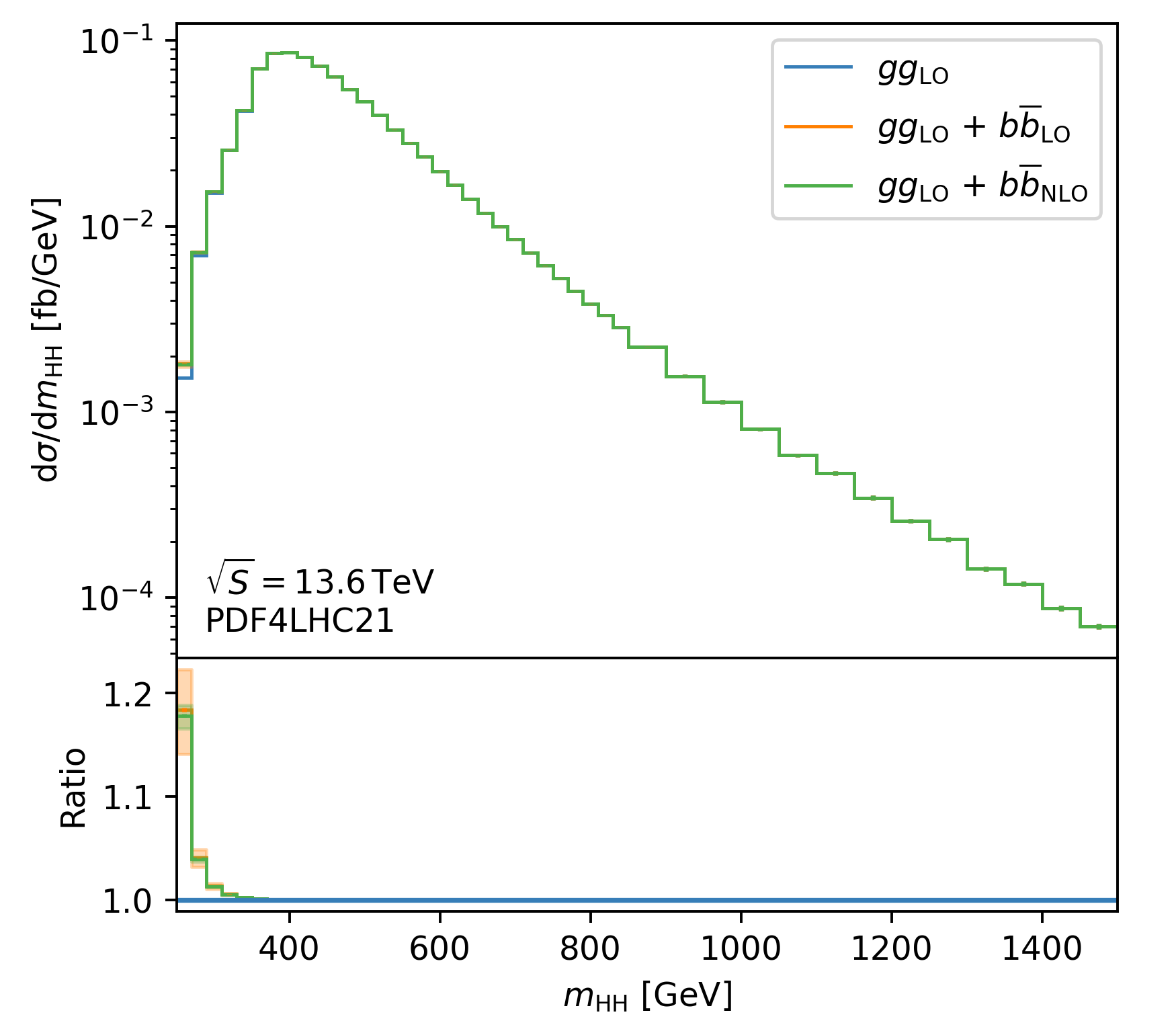}
    \includegraphics[width=0.48\linewidth]{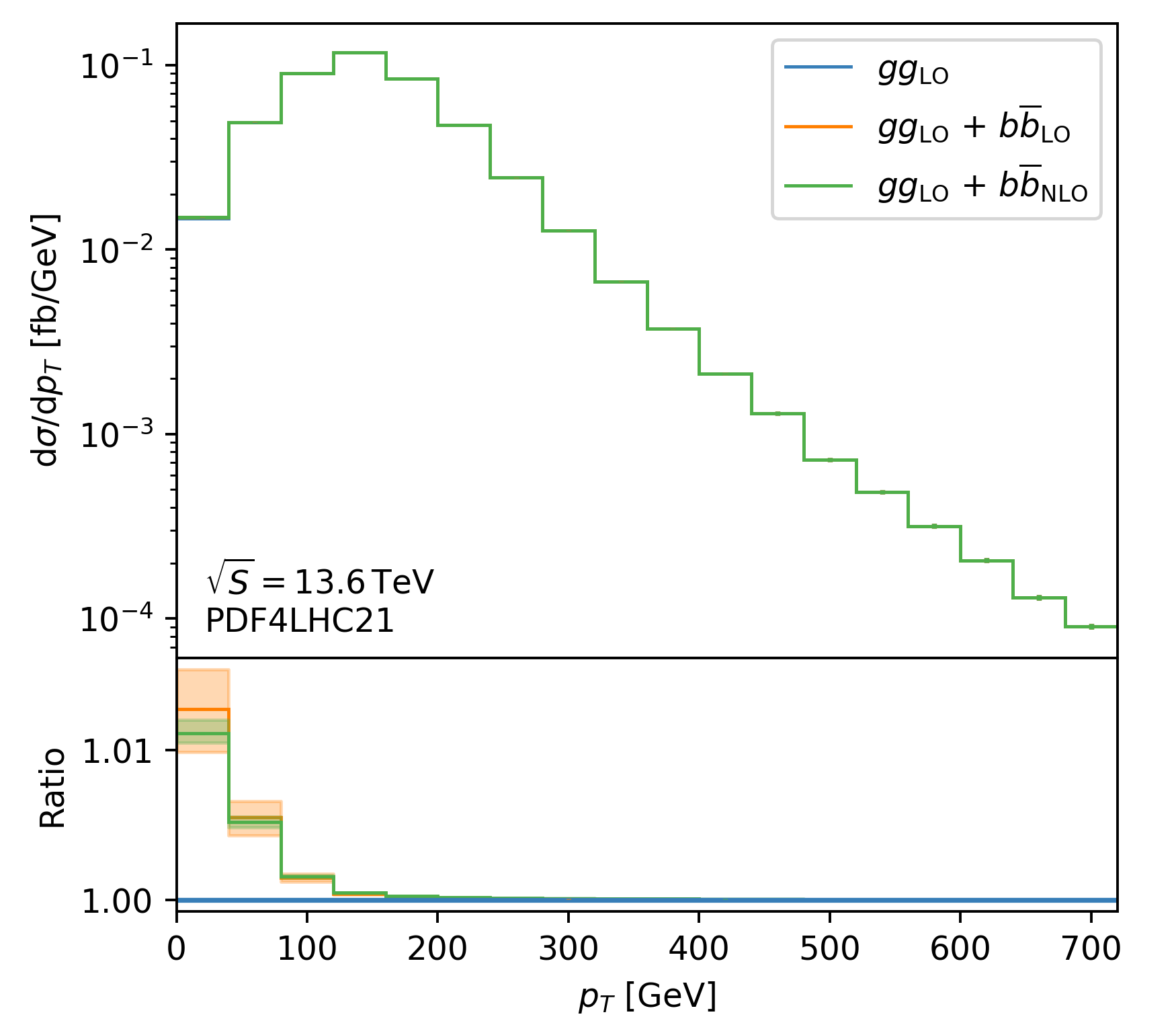}
    \caption{Invariant mass distribution of the Higgs pair (left) and transverse momentum distribution of a single Higgs boson (right) at LO (orange) and NLO (green) in the bottom quark annihilation channel, plotted against the LO gluon fusion channel (blue). The uncertainty bands represent the scale uncertainty obtained from a 7-point scale variation.}
    \label{fig:bbHH-dists}
\end{figure}

In Figure~\ref{fig:bbHH-dists} we present the invariant mass distribution of the Higgs-boson pair (left) and the transverse momentum distribution of one of the Higgs bosons (right) for proton-proton collisions at a centre-of-mass energy of $13.6\,\mathrm{TeV}$. The green (orange) line shows bottom quark annihilation at NLO (LO) plotted against the LO gluon fusion channel in blue. The error bands represent the scale uncertainty, obtained from a 7-point scale variation around the central value $\renscale = \facscale = m_{HH}/2$.

The left-hand side of Figure~\ref{fig:bbHH-dists} shows the invariant mass distribution of the Higgs boson pair. The inclusion of bottom quark annihilation at LO leads to an enhancement of up to $18\%$ in the left-most bin with respect to the LO gluon fusion. The further inclusion of NLO virtual corrections and real emissions reduces this effect slightly. Towards higher energies the LO gluon fusion becomes larger, causing the relative effect of bottom quark annihilation going to zero. 

The transverse momentum distribution of one of the Higgs bosons is shown on the right-hand side of Figure~\ref{fig:bbHH-dists}. Here, the effect is also largest for small $\pth$-values, reaching a correction up to $1.1\%$ at LO and $1.0\%$ at NLO.

We observe for both distributions that the NLO scale uncertainty band is smaller by about a factor of two and fully covered by the LO uncertainty band, indicating a good convergence of the perturbative series. 

\newpage

\section{Combined threshold effects}
\label{sec:threshold}
\begin{figure}[tb]
    \centering
    \includegraphics[width=0.98\linewidth]{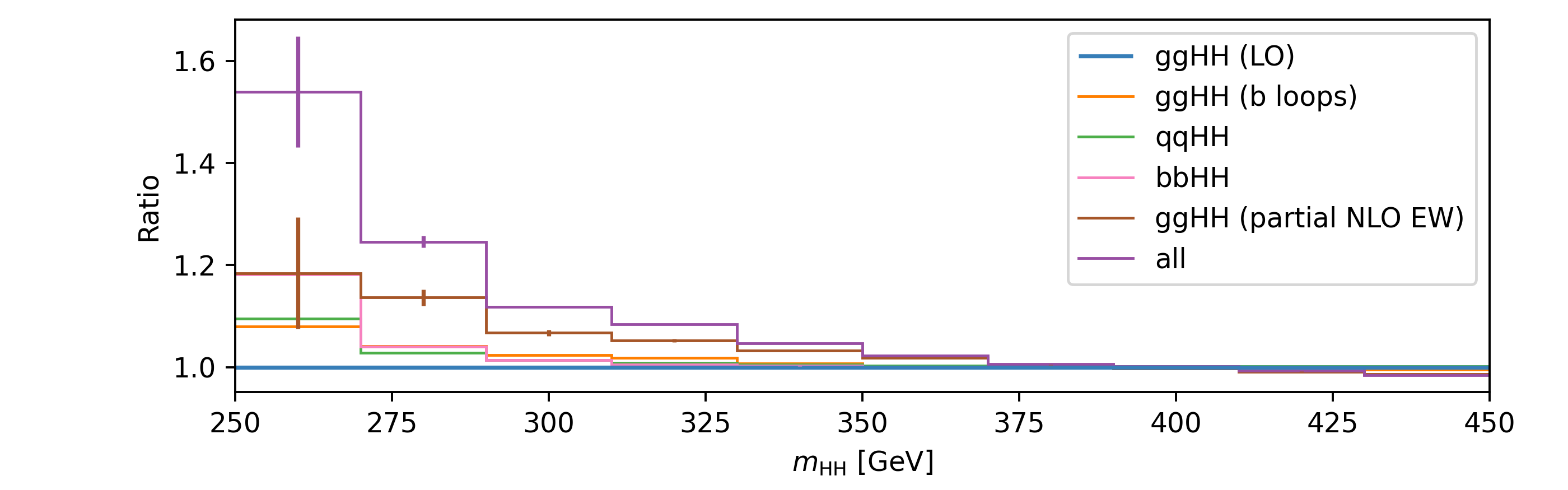}
    \caption{Comparison of the threshold enhancements from the different contributions to Higgs-boson pair production, shown relative to the LO gluon-fusion prediction. The brown curve shows partial NLO EW corrections to gluon fusion, consisting of the top-Yukawa, self-coupling and light quark contributions, which reproduces the threshold behaviour of the full NLO EW corrections. The other curves show the threshold effects from bottom quark annihilation (pink), the quark--antiquark channel at NLO QCD (green), and massive bottom quark loop effects in LO gluon fusion (orange). Their combined effect is shown in purple.}
    \label{fig:threshold}
\end{figure}

Bottom quark effects, both in loops and as external particles, provide an example of subleading corrections that have a small impact on the total cross section but can become significantly enhanced in specific kinematic regions. Similar threshold effects are observed in other contributions to Higgs boson pair production. In particular, such enhancements arise in the electroweak production of a Higgs boson pair through the quark--antiquark channel, computed up to NLO QCD in Ref.~\cite{Bonetti:2026cih}, as well as in the NLO EW corrections to gluon fusion, first computed in Ref.~\cite{Bi:2023bnq}. 

Figure~\ref{fig:threshold} compares these individual threshold effects and their combined impact. All contributions are shown relative to the LO gluon fusion prediction (blue). The largest enhancement is due to partial NLO EW corrections to gluon fusion (brown), which are a combination of the top-Yukawa and Higgs self-coupling contributions of Ref.~\cite{Heinrich:2024dnz} with the light-quark contribution of Ref.~\cite{Bonetti:2025vfd}.\footnote{Related subsets or limits of the NLO EW corrections were also computed in Ref.~\cite{Muhlleitner:2022ijf,Davies:2022ram,Davies:2023npk,Bhattacharya:2025wnr}.} A comparison with Ref.~\cite{Bi:2023bnq} shows that these contributions already reproduce the correct threshold behaviour of the full NLO EW corrections within the statistical uncertainties. 

This is followed by bottom quark annihilation (pink), the quark--antiquark channel at NLO QCD (green), and the contribution from massive bottom quark loops to LO gluon fusion (orange). Combining these contributions leads to the threshold behaviour shown in purple, with a maximal enhancement of approximately $\mathcal{O}(54\%)$ in the leftmost bin.

\section{Conclusions}
\label{sec:conclusion}
We have investigated the effects of bottom quarks on Higgs boson pair production in two different channels: For the gluon fusion channel, we have studied how bottom quark loops affect the Higgs boson pair invariant mass distribution using the on-shell scheme or the $\msbar$ scheme for the bottom quark mass, in combination with different renormalisation schemes for the top quark mass.
For the bottom quark annihilation channel, we have calculated the NLO QCD corrections and studied their impact on the $\mhh$ and $\pth$ distributions.
We have implemented the NLO corrections for $b\bar{b}\to HH$ in the code \texttt{bbHH} within the \POWHEGBOXVV{} framework, which will be publicly available at
\begin{center}
\url{https://gitlab.com/hep-mirrors/powheg-box-v2-user-processes.git}
\end{center}

For the gluon fusion channel, we have found that, at leading order, with the bottom quark mass in the $\msbar$ scheme and a bin size of $25\,\mathrm{GeV}$, the $\mhh$ distribution near the Higgs pair production threshold increases by approximately $6\,\%$ when taking massive bottom quark loops into account. With the bottom quark mass in the on-shell scheme, the differential cross section near the Higgs pair production threshold increases by approximately $12\,\%$.

In addition, we have compared the four- and five-flavour schemes, finding that the differences between these schemes are negligible.
We have also studied the scale dependence including variations of the $\msbar$ mass scale $\mu_m$ and the influence of the order in $\alpha_s$ that is used in the mass conversion and the renormalisation group evolution of the quark masses. We found that the scale uncertainties can be artificially small when using $\mu_{m_t}=
\mu_r$, compared to a 21-point scale variation. 
The differences due to the order in $\alpha_s$ that is used in the mass conversion are usually small, but can be of order 10\% around the top quark pair production threshold if both the top- and the bottom quark are renormalised in the $\msbar$ scheme.

For the bottom quark annihilation channel, we have found that the inclusion of bottom quark annihilation at LO leads to an enhancement of up to $18\%$ at low-$\mhh$ with respect to the LO gluon fusion, while the peak region and the tail of the $\mhh$ distribution are not affected. The NLO corrections to the $b\bar{b}$ channel reduce this effect slightly and lead to a reduction of the scale uncertainties by about a factor of two.

In summary, the $\mhh$ region close to the Higgs boson pair production threshold is very sensitive to various effects: 
\begin{itemize}
    \item bottom quark loops in the gluon fusion channel
    \item electroweak corrections to the gluon fusion channel 
    \item contributions from bottom quark annihilation 
    \item electroweak contributions in the quark channel, i.e. $q\bar{q}\to HH$ calculated in Ref.~\cite{Bonetti:2026cih}, including NLO QCD corrections.
\end{itemize}
All these corrections lead to an enhancement at low $\mhh$ values, which is interesting because this region is also quite sensitive to non-SM values of the trilinear Higgs coupling.

\section*{Acknowledgements}

This research is supported in parts by the Deutsche Forschungsgemeinschaft (DFG, German Research Foundation) under grant 396021762 - TRR 257.

\bibliographystyle{SciPost_bibstyle}
\bibliography{HH_mb}

@article{Alioli:2010xd,
    author = "Alioli, Simone and Nason, Paolo and Oleari, Carlo and Re, Emanuele",
    title = "{A general framework for implementing NLO calculations in shower Monte Carlo programs: the POWHEG BOX}",
    eprint = "1002.2581",
    archivePrefix = "arXiv",
    primaryClass = "hep-ph",
    reportNumber = "DESY-10-018, SFB-CPP-10-22, IPPP-10-11, DCPT-10-22",
    doi = "10.1007/JHEP06(2010)043",
    journal = "JHEP",
    volume = "06",
    pages = "043",
    year = "2010"
}

@article{Buchalla:2018yce,
    author = "Buchalla, G. and Capozi, M. and Celis, A. and Heinrich, G. and Scyboz, L.",
    title = "{Higgs boson pair production in non-linear Effective Field Theory with full $m_t$-dependence at NLO QCD}",
    eprint = "1806.05162",
    archivePrefix = "arXiv",
    primaryClass = "hep-ph",
    reportNumber = "LMU-ASC 34/18, MPP-2018-127, LMU-ASC-34-18",
    doi = "10.1007/JHEP09(2018)057",
    journal = "JHEP",
    volume = "09",
    pages = "057",
    year = "2018",
    note = "[Erratum: JHEP 06, 094 (2025)]"
}

@article{Heinrich:2020ckp,
    author = "Heinrich, Gudrun and Jones, Stephen P. and Kerner, Matthias and Scyboz, Ludovic",
    title = "{A non-linear EFT description of $gg\to HH$ at NLO interfaced to POWHEG}",
    eprint = "2006.16877",
    archivePrefix = "arXiv",
    primaryClass = "hep-ph",
    reportNumber = "CERN-TH-2020-106, KA-TP-04-2020, OUTP-20-06P, P3H-20-026,
  ZU-TH-21/20, P3H-20-026, ZU-TH-21/20",
    doi = "10.1007/JHEP10(2020)021",
    journal = "JHEP",
    volume = "10",
    pages = "021",
    year = "2020"
}

@article{Baglio:2018lrj,
    author = {Baglio, Julien and Campanario, Francisco and Glaus, Seraina and M{\"u}hlleitner, Margarete and Spira, Michael and Streicher, Juraj},
    title = "{Gluon fusion into Higgs pairs at NLO QCD and the top mass scheme}",
    eprint = "1811.05692",
    archivePrefix = "arXiv",
    primaryClass = "hep-ph",
    reportNumber = "FTUV-18-1113, IFIC/18-55, KA-TP-32-2018, PSI-PR-18-14",
    doi = "10.1140/epjc/s10052-019-6973-3",
    journal = "Eur. Phys. J. C",
    volume = "79",
    number = "6",
    pages = "459",
    year = "2019"
}

@article{Davies:2025qjr,
    author = {Davies, Joshua and Sch{\"o}nwald, Kay and Steinhauser, Matthias and Stremmer, Daniel},
    title = "{{{\textbackslash}tt ggxy}: a flexible library to compute gluon-induced cross sections}",
    eprint = "2506.04323",
    archivePrefix = "arXiv",
    primaryClass = "hep-ph",
    reportNumber = "LTH 1406, P3H-25-035, TTP25-017, ZU-TH40/25",
    month = "6",
    year = "2025"
}

@article{Bagnaschi:2023rbx,
    author = {Bagnaschi, Emanuele and Degrassi, Giuseppe and Gr{\"o}ber, Ramona},
    title = "{Higgs boson pair production at NLO in the POWHEG approach and the top quark mass uncertainties}",
    eprint = "2309.10525",
    archivePrefix = "arXiv",
    primaryClass = "hep-ph",
    reportNumber = "CERN-TH-2023-131",
    doi = "10.1140/epjc/s10052-023-12238-8",
    journal = "Eur. Phys. J. C",
    volume = "83",
    number = "11",
    pages = "1054",
    year = "2023"
}

@article{Glover:1987nx,
    author = "Glover, E. W. Nigel and van der Bij, J. J.",
    title = "{Higgs boson pair production via gluon fusion}",
    reportNumber = "CERN-TH-4934/87",
    doi = "10.1016/0550-3213(88)90083-1",
    journal = "Nucl. Phys. B",
    volume = "309",
    pages = "282--294",
    year = "1988"
}

@article{Nason:2004rx,
    author = "Nason, Paolo",
    title = "{A new method for combining NLO QCD with shower Monte Carlo algorithms}",
    eprint = "hep-ph/0409146",
    archivePrefix = "arXiv",
    reportNumber = "BICOCCA-FT-04-11",
    doi = "10.1088/1126-6708/2004/11/040",
    journal = "JHEP",
    volume = "11",
    pages = "040",
    year = "2004"
}

@article{Frixione:2007vw,
    author = "Frixione, Stefano and Nason, Paolo and Oleari, Carlo",
    title = "{Matching NLO QCD computations with parton shower simulations: the POWHEG method}",
    eprint = "0709.2092",
    archivePrefix = "arXiv",
    primaryClass = "hep-ph",
    reportNumber = "BICOCCA-FT-07-9, GEF-TH-21-2007",
    doi = "10.1088/1126-6708/2007/11/070",
    journal = "JHEP",
    volume = "11",
    pages = "070",
    year = "2007"
}

@article{Buckley:2014ana,
    author = {Buckley, Andy and Ferrando, James and Lloyd, Stephen and Nordstr{\"o}m, Karl and Page, Ben and R{\"u}fenacht, Martin and Sch{\"o}nherr, Marek and Watt, Graeme},
    title = "{LHAPDF6: parton density access in the LHC precision era}",
    eprint = "1412.7420",
    archivePrefix = "arXiv",
    primaryClass = "hep-ph",
    reportNumber = "GLAS-PPE-2014-05, MCNET-14-29, IPPP-14-111, DCPT-14-222",
    doi = "10.1140/epjc/s10052-015-3318-8",
    journal = "Eur. Phys. J. C",
    volume = "75",
    pages = "132",
    year = "2015"
}

@article{Heinrich:2017kxx,
    author = "Heinrich, G. and Jones, S. P. and Kerner, M. and Luisoni, G. and Vryonidou, E.",
    title = "{NLO predictions for Higgs boson pair production with full top quark mass dependence matched to parton showers}",
    eprint = "1703.09252",
    archivePrefix = "arXiv",
    primaryClass = "hep-ph",
    reportNumber = "CERN-TH-2017-069, MPP-2017-41, NIKHEF-2017-020",
    doi = "10.1007/JHEP08(2017)088",
    journal = "JHEP",
    volume = "08",
    pages = "088",
    year = "2017"
}

@article{Melnikov:2000qh,
    author = "Melnikov, Kirill and Ritbergen, Timo van",
    title = "{The Three loop relation between the MS-bar and the pole quark masses}",
    eprint = "hep-ph/9912391",
    archivePrefix = "arXiv",
    reportNumber = "SLAC-PUB-8321, TTP-99-51",
    doi = "10.1016/S0370-2693(00)00507-4",
    journal = "Phys. Lett. B",
    volume = "482",
    pages = "99--108",
    year = "2000"
}

@article{Czakon:2024ywb,
    author = "Czakon, Micha{\l} and Eschment, Felix and Niggetiedt, Marco and Poncelet, Rene and Schellenberger, Tom",
    title = "{Quark mass effects in Higgs production}",
    eprint = "2407.12413",
    archivePrefix = "arXiv",
    primaryClass = "hep-ph",
    doi = "10.1007/JHEP10(2024)210",
    journal = "JHEP",
    volume = "10",
    pages = "210",
    year = "2024"
}

@article{Harlander:2003ai,
    author = "Harlander, Robert V. and Kilgore, William B.",
    title = "{Higgs boson production in bottom quark fusion at next-to-next-to leading order}",
    eprint = "hep-ph/0304035",
    archivePrefix = "arXiv",
    reportNumber = "BNL-HET-03-4, CERN-TH-2003-067",
    doi = "10.1103/PhysRevD.68.013001",
    journal = "Phys. Rev. D",
    volume = "68",
    pages = "013001",
    year = "2003"
}

@article{Dawson:2006dm,
    author = "Dawson, Sally and Kao, Chung and Wang, Yili and Williams, Peter",
    title = "{QCD Corrections to Higgs Pair Production in Bottom Quark Fusion}",
    eprint = "hep-ph/0610284",
    archivePrefix = "arXiv",
    reportNumber = "FERMILAB-PUB-06-378-T",
    doi = "10.1103/PhysRevD.75.013007",
    journal = "Phys. Rev. D",
    volume = "75",
    pages = "013007",
    year = "2007"
}

@article{AH:2018tzg,
    author = "A H, Ajjath and Banerjee, Pulak and Chakraborty, Amlan and Dhani, Prasanna K. and Mukherjee, Pooja and Rana, Narayan and Ravindran, V.",
    title = "{Higgs pair production from bottom quark annihilation to NNLO in QCD}",
    eprint = "1811.01853",
    archivePrefix = "arXiv",
    primaryClass = "hep-ph",
    reportNumber = "IMSc/2018/11/09, DESY-18-166, TIF-UNIMI-2018-9",
    doi = "10.1007/JHEP05(2019)030",
    journal = "JHEP",
    volume = "05",
    pages = "030",
    year = "2019"
}

@article{Ahmed:2021hrf,
    author = "Ahmed, Taushif and Ravindran, V. and Sankar, Aparna and Tiwari, Surabhi",
    title = "{Two-loop amplitudes for di-Higgs and di-pseudo-Higgs productions through quark annihilation in QCD}",
    eprint = "2110.11476",
    archivePrefix = "arXiv",
    primaryClass = "hep-ph",
    reportNumber = "IMSc/2021/10/08",
    doi = "10.1007/JHEP01(2022)189",
    journal = "JHEP",
    volume = "01",
    pages = "189",
    year = "2022"
}

@article{Braun:2025afl,
    author = {Braun, Jens and Campillo Aveleira, Benjamin and Heinrich, Gudrun and H{\"o}fer, Marius and Jones, Stephen P. and Kerner, Matthias and Lang, Jannis and Magerya, Vitaly},
    title = "{One-loop calculations in effective field theories with GoSam-3.0}",
    eprint = "2507.23549",
    archivePrefix = "arXiv",
    primaryClass = "hep-ph",
    reportNumber = "KA-TP-21-2025, P3H-25-051, IPPP/25/50, CERN-TH-2025-134",
    doi = "10.21468/SciPostPhysCodeb.62",
    journal = "SciPost Phys. Codeb.",
    volume = "62",
    pages = "1",
    year = "2026"
}

@article{Jaskiewicz:2024xkd,
    author = "Jaskiewicz, Sebastian and Jones, Stephen and Szafron, Robert and Ulrich, Yannick",
    title = "{The structure of quark mass corrections in the gg {\textrightarrow} HH amplitude at high-energy}",
    eprint = "2501.00587",
    archivePrefix = "arXiv",
    primaryClass = "hep-ph",
    reportNumber = "IPPP/24/83, LTH 1389",
    doi = "10.1007/JHEP09(2025)015",
    journal = "JHEP",
    volume = "09",
    pages = "015",
    year = "2025"
}

@article{Hu:2025aeo,
    author = "Hu, Zhenghong and Liu, Tao and Yang, Jin Min",
    title = "{The gg {\textrightarrow} HH amplitude induced by bottom quarks at two-loop level: planar master integrals}",
    eprint = "2503.10051",
    archivePrefix = "arXiv",
    primaryClass = "hep-ph",
    doi = "10.1007/JHEP09(2025)132",
    journal = "JHEP",
    volume = "09",
    pages = "132",
    year = "2025"
}

@article{Davies:2025ghl,
    author = {Davies, Joshua and Sch{\"o}nwald, Kay and Steinhauser, Matthias},
    title = "{Three-loop large-N$_{c}$ virtual corrections to gg {\textrightarrow} HH in the forward limit}",
    eprint = "2503.17449",
    archivePrefix = "arXiv",
    primaryClass = "hep-ph",
    reportNumber = "LTH 1397, P3H-25-020, TTP25-007, ZU-TH 19/25",
    doi = "10.1007/JHEP08(2025)192",
    journal = "JHEP",
    volume = "08",
    pages = "192",
    year = "2025"
}

@article{PDF4LHCWorkingGroup:2022cjn,
    author = "Ball, Richard D. and others",
    collaboration = "PDF4LHC Working Group",
    title = "{The PDF4LHC21 combination of global PDF fits for the LHC Run III}",
    eprint = "2203.05506",
    archivePrefix = "arXiv",
    primaryClass = "hep-ph",
    reportNumber = "Edinburgh 2021/31, FERMILAB-PUB-22-121-QIS-SCD-T, MSUHEP-22-010, SMU-HEP-22-01, Nikhef 2021-033",
    doi = "10.1088/1361-6471/ac7216",
    journal = "J. Phys. G",
    volume = "49",
    number = "8",
    pages = "080501",
    year = "2022"
}

@article{Harland-Lang:2014zoa,
    author = "Harland-Lang, L. A. and Martin, A. D. and Motylinski, P. and Thorne, R. S.",
    title = "{Parton distributions in the LHC era: MMHT 2014 PDFs}",
    eprint = "1412.3989",
    archivePrefix = "arXiv",
    primaryClass = "hep-ph",
    reportNumber = "LCTS-2014-47, IPPP-14-97, DCPT-14-194",
    doi = "10.1140/epjc/s10052-015-3397-6",
    journal = "Eur. Phys. J. C",
    volume = "75",
    number = "5",
    pages = "204",
    year = "2015"
}

@article{ATLAS:2025eii,
    author = "Aad, Georges and others",
    collaboration = "ATLAS, CMS",
    title = "{Highlights of the HL-LHC physics projections by ATLAS and CMS}",
    eprint = "2504.00672",
    archivePrefix = "arXiv",
    primaryClass = "hep-ex",
    reportNumber = "ATL-PHYS-PUB-2025-018, CMS-HIG-25-002",
    month = "4",
    year = "2025"
}

@article{Chen:2026zmi,
    author = "Chen, Xuan and Dai, Yuesheng and Li, Hai Tao and Li, Shi-Yuan and Shao, Hua-Sheng and Wang, Jian",
    title = "{Fully differential Higgs boson pair production at N$^{3}$LO with top quark mass effects}",
    eprint = "2601.19990",
    archivePrefix = "arXiv",
    primaryClass = "hep-ph",
    doi = "10.1007/JHEP06(2026)005",
    journal = "JHEP",
    volume = "06",
    pages = "005",
    year = "2026"
}

@article{Bonetti:2026cih,
    author = "Bonetti, Marco and Heinrich, Gudrun and Rendler, Philipp and Torres Bobadilla, William J.",
    title = "{NLO QCD corrections to the electroweak production of a Higgs boson pair in the quark-antiquark channel}",
    eprint = "2601.16924",
    archivePrefix = "arXiv",
    primaryClass = "hep-ph",
    reportNumber = "KA-TP-04-2026, P3H-26-008",
    doi = "10.1007/JHEP04(2026)131",
    journal = "JHEP",
    volume = "04",
    pages = "131",
    year = "2026"
}

@article{Gross:1973id,
    author = "Gross, David J. and Wilczek, Frank",
    title = "{Ultraviolet Behavior of Nonabelian Gauge Theories}",
    doi = "10.1103/PhysRevLett.30.1343",
    journal = "Phys. Rev. Lett.",
    volume = "30",
    pages = "1343--1346",
    year = "1973"
}

@article{Politzer:1973fx,
    author = "Politzer, H. David",
    title = "{Reliable Perturbative Results for Strong Interactions?}",
    doi = "10.1103/PhysRevLett.30.1346",
    journal = "Phys. Rev. Lett.",
    volume = "30",
    pages = "1346--1349",
    year = "1973"
}

@article{Jones:1974mm,
    author = "Jones, D. R. T.",
    title = "{Two Loop Diagrams in Yang-Mills Theory}",
    reportNumber = "OXFORD-TP-10-74",
    doi = "10.1016/0550-3213(74)90093-5",
    journal = "Nucl. Phys. B",
    volume = "75",
    pages = "531",
    year = "1974"
}

@article{Caswell:1974gg,
    author = "Caswell, William E.",
    title = "{Asymptotic Behavior of Nonabelian Gauge Theories to Two Loop Order}",
    reportNumber = "PRINT-74-1058 (PRINCETON)",
    doi = "10.1103/PhysRevLett.33.244",
    journal = "Phys. Rev. Lett.",
    volume = "33",
    pages = "244",
    year = "1974"
}

@article{Tarrach:1980up,
    author = "Tarrach, R.",
    title = "{The Pole Mass in Perturbative QCD}",
    reportNumber = "CPT-80/P-1215",
    doi = "10.1016/0550-3213(81)90140-1",
    journal = "Nucl. Phys. B",
    volume = "183",
    pages = "384--396",
    year = "1981"
}

@article{Nachtmann:1981zg,
    author = "Nachtmann, O. and Wetzel, W.",
    title = "{The Beta Function for Effective Quark Masses to Two Loops in {QCD}}",
    reportNumber = "HD-THEP-81-1",
    doi = "10.1016/0550-3213(81)90278-9",
    journal = "Nucl. Phys. B",
    volume = "187",
    pages = "333--342",
    year = "1981"
}

@article{Tarasov:1980au,
    author = "Tarasov, O. V. and Vladimirov, A. A. and Zharkov, A. Yu.",
    title = "{The Gell-Mann-Low Function of QCD in the Three Loop Approximation}",
    doi = "10.1016/0370-2693(80)90358-5",
    journal = "Phys. Lett. B",
    volume = "93",
    pages = "429--432",
    year = "1980"
}

@article{Larin:1993tq,
    author = "Larin, S. A.",
    title = "{The Renormalization of the axial anomaly in dimensional regularization}",
    eprint = "hep-ph/9302240",
    archivePrefix = "arXiv",
    reportNumber = "NIKHEF-H-92-18",
    doi = "10.1016/0370-2693(93)90053-K",
    journal = "Phys. Lett. B",
    volume = "303",
    pages = "113--118",
    year = "1993"
}

@article{Bi:2023bnq,
    author = "Bi, Huan-Yu and Huang, Li-Hong and Huang, Rui-Jun and Ma, Yan-Qing and Yu, Huai-Min",
    title = "{Electroweak Corrections to Double Higgs Production at the LHC}",
    eprint = "2311.16963",
    archivePrefix = "arXiv",
    primaryClass = "hep-ph",
    doi = "10.22323/1.478.0120",
    journal = "Phys. Rev. Lett.",
    volume = "132",
    number = "23",
    pages = "231802",
    year = "2024"
}

@article{Heinrich:2024dnz,
    author = "Heinrich, Gudrun and Jones, Stephen and Kerner, Matthias and Stone, Thomas and Vestner, Augustin",
    title = "{Electroweak corrections to Higgs boson pair production: the top-Yukawa and self-coupling contributions}",
    eprint = "2407.04653",
    archivePrefix = "arXiv",
    primaryClass = "hep-ph",
    reportNumber = "KA-TP-11-2024, IPPP/24/43, P3H-24-046",
    doi = "10.1007/JHEP11(2024)040",
    journal = "JHEP",
    volume = "11",
    pages = "040",
    year = "2024"
}

@article{Bonetti:2025vfd,
    author = "Bonetti, Marco and Rendler, Philipp and Torres Bobadilla, William J.",
    title = "{Two-loop light-quark Electroweak corrections to Higgs boson pair production in gluon fusion}",
    eprint = "2503.16620",
    archivePrefix = "arXiv",
    primaryClass = "hep-ph",
    reportNumber = "KA-TP-06-2025, P3H-25-018",
    doi = "10.1007/JHEP07(2025)024",
    journal = "JHEP",
    volume = "07",
    pages = "024",
    year = "2025"
}

@article{Muhlleitner:2022ijf,
    author = {M{\"u}hlleitner, Margarete and Schlenk, Johannes and Spira, Michael},
    title = "{Top-Yukawa-induced corrections to Higgs pair production}",
    eprint = "2207.02524",
    archivePrefix = "arXiv",
    primaryClass = "hep-ph",
    reportNumber = "KA--TP--18--2022, PSI--PR--22--20",
    doi = "10.1007/JHEP10(2022)185",
    journal = "JHEP",
    volume = "10",
    pages = "185",
    year = "2022"
}

@article{Bhattacharya:2025wnr,
    author = {Bhattacharya, Arunima and Campanario, Francisco and Carlotti, Sauro and Chang, Jamie and Mazzitelli, Javier and M{\"u}hlleitner, Milada and Ronca, Jonathan and Spira, Michael},
    title = "{Higgs-pair production via gluon fusion: top-Yukawa-and light-quark-induced electroweak corrections}",
    eprint = "2512.14823",
    archivePrefix = "arXiv",
    primaryClass = "hep-ph",
    reportNumber = "KA-TP-35-2025",
    doi = "10.1007/JHEP07(2026)057",
    journal = "JHEP",
    volume = "07",
    pages = "057",
    year = "2026"
}

@article{Davies:2022ram,
    author = {Davies, Joshua and Mishima, Go and Sch{\"o}nwald, Kay and Steinhauser, Matthias and Zhang, Hantian},
    title = "{Higgs boson contribution to the leading two-loop Yukawa corrections to gg {\textrightarrow} HH}",
    eprint = "2207.02587",
    archivePrefix = "arXiv",
    primaryClass = "hep-ph",
    reportNumber = "P3H-22-063, TTP22-041, TU-1160",
    doi = "10.1007/JHEP08(2022)259",
    journal = "JHEP",
    volume = "08",
    pages = "259",
    year = "2022"
}

@article{Davies:2023npk,
    author = {Davies, Joshua and Sch{\"o}nwald, Kay and Steinhauser, Matthias and Zhang, Hantian},
    title = "{Next-to-leading order electroweak corrections to $gg \to HH$ and $gg \to gH$ in the large-$m_t$ limit}",
    eprint = "2308.01355",
    archivePrefix = "arXiv",
    primaryClass = "hep-ph",
    reportNumber = "P3H-23-051, TTP23-029, ZU-TH 40/23",
    doi = "10.1007/JHEP10(2023)033",
    journal = "JHEP",
    volume = "10",
    pages = "033",
    year = "2023"
}

@article{Alonso:2012px,
    author = "Alonso, R. and Gavela, M. B. and Merlo, L. and Rigolin, S. and Yepes, J.",
    title = "{The Effective Chiral Lagrangian for a Light Dynamical ''Higgs Particle''}",
    eprint = "1212.3305",
    archivePrefix = "arXiv",
    primaryClass = "hep-ph",
    reportNumber = "FTUAM-12-115, IFT-UAM-CSIC-12-113, CERN-PH-TH-2012-335, DFPD-2012-TH-23",
    doi = "10.1016/j.physletb.2013.04.037",
    journal = "Phys. Lett. B",
    volume = "722",
    pages = "330--335",
    year = "2013",
    note = "[Erratum: Phys.Lett.B 726, 926 (2013)]"
}

@article{Buchalla:2013rka,
    author = "Buchalla, Gerhard and Cat{\`a}, Oscar and Krause, Claudius",
    title = "{Complete Electroweak Chiral Lagrangian with a Light Higgs at NLO}",
    eprint = "1307.5017",
    archivePrefix = "arXiv",
    primaryClass = "hep-ph",
    reportNumber = "LMU-ASC-42-13, LMU-ASC{\textasciitilde}42-13",
    doi = "10.1016/j.nuclphysb.2014.01.018",
    journal = "Nucl. Phys. B",
    volume = "880",
    pages = "552--573",
    year = "2014",
    note = "[Erratum: Nucl.Phys.B 913, 475--478 (2016)]"
}

\end{document}